\documentclass[preprint,aps]{revtex4-2}
\usepackage{mathrsfs}
\usepackage{graphicx}
\usepackage{multirow}
\usepackage{makecell}
\usepackage{booktabs}
\usepackage{hyperref}
\hypersetup{
    colorlinks,
    citecolor=black,
    filecolor=black,
    linkcolor=black,
    urlcolor=black
}
\usepackage{color}
\usepackage{bm}

\begin{document}

\title{Topological Weyl Altermagnetism in CrSb}

\author{Cong Li$^{1,*,\sharp}$, Mengli Hu$^{2,*}$, Zhilin Li$^{3,*}$, Yang Wang$^{1}$, Wanyu Chen$^{1}$, Balasubramanian Thiagarajan$^{4}$, Mats Leandersson$^{4}$, Craig Polley$^{4}$, Timur Kim$^{5}$, Hui Liu$^{6,\sharp}$, Cosma Fulga$^{2,7}$, Maia G. Vergniory$^{8,9}$, Oleg Janson$^{2}$, Oscar Tjernberg$^{1,\sharp}$, Jeroen van den Brink$^{2,7,\sharp}$
}

\affiliation{
\\$^{1}$Department of Applied Physics, KTH Royal Institute of Technology, Stockholm 11419, Sweden
\\$^{2}$Leibniz Institute for Solid State and Materials Research, IFW Dresden, Helmholtzstraße 20, 01069 Dresden, Germany
\\$^{3}$Beijing National Laboratory for Condensed Matter Physics, Institute of Physics, Chinese Academy of Sciences, Beijing 100190, China
\\$^{4}$MAX IV Laboratory, Lund University, 22100 Lund, Sweden
\\$^{5}$Diamond Light Source, Harwell Campus, Didcot, OX11 0DE, United Kingdom
\\$^{6}$Department of Physics, Stockholm University, AlbaNova University Center, 10691 Stockholm, Sweden
\\$^{7}$Würzburg-Dresden Cluster of Excellence Ct.qmat, Technische Universitat Dresden, 01062, Dresden, Germany
\\$^{8}$Max Planck Institute for Chemical Physics of Solids, 01187 Dresden, Germany
\\$^{9}$Donostia International Physics Center, 20018 Donostia–San Sebastian, Spain
\\$^{*}$These authors contributed equally to the present work.
\\$^{\sharp}$Corresponding authors: conli@kth.se, hui.liu@fysik.su.se, oscar@kth.se, j.van.den.brink@ifw-dresden.de
}


\maketitle


\begin{center}
{\bf Abstract}
\end{center}

{\bf 
Altermagnets constitute a novel, third fundamental class of collinear magnetic ordered materials, alongside with ferro- and antiferromagnets \cite{LSmejkal_PRX2022_1_TJungwirth,LSmejkal_PRX2022_2_TJungwirth}. 
They share with conventional antiferromagnets the feature of a vanishing net magnetization. 
At the same time they show a spin-splitting of electronic bands, just as in ferromagnets, caused by the atomic exchange interaction\cite{LDYuan_PRB2020_AZunger,SHayami_JPSJ2019_HKusunose,Yaqian_Guo_Material_today_physics}. 
On the other hand, topology has recently revolutionized our understanding of condensed matter physics, introducing new phases of matter classified by intrinsic topological order\cite{NPArmitage_RMP2018_AVishwanath,MGVergniory_Nature2019_ZJWang,YFXu_Nature2020_BABernevig}. 
Here we connect the worlds of altermagnetism and topology, showing that the electronic structure of the altermagnet CrSb is topological and hosts a novel Weyl semimetallic state. 
Using high-resolution and spin angle-resolved photoemission spectroscopy, we observe a large momentum-dependent spin-splitting in CrSb, reaching up to 1 eV, that
induces altermagnetic Weyl nodes with an associated magnetic quantum number. 
At the surface we observe their spin-polarized topological Fermi-arcs. 
This establishes that in altermagnets the large energy scale intrinsic to the spin-splitting  -- orders of magnitude larger than the relativistic spin-orbit coupling -- creates its own realm of robust electronic topology.
}

The recently established new class of altermagnetic materials is characterized by the presence of particular alternating opposite-spin sublattices. They defy conventional magnetic classifications by breaking time-reversal symmetry while simultaneously maintaining a zero net magnetization enforced by spin-lattice symmetry\cite{LSmejkal_PRX2022_1_TJungwirth,LSmejkal_PRX2022_2_TJungwirth}. Precisely this intrinsic trait allows altermagnets to combine advantageous features of both ferromagnets and antiferromagnets\cite{TJungwirth_NN2016_JWunderlich,VBaltz_RMP2018_YTserkovnyak}, including spin-transport properties and compatibility with a diverse range of materials\cite{KPKluczyk_Arxiv2023_MGBorysiewicz,TSato_Arxiv2023_JVDBrink,ZXFeng_NE2022_ZQLiu,HBai_PRL2023_CSong,MLeiviska_Arxiv2024_VBaltz,TUrata_Arxiv2024_HIkuta,HMa_NC2021_JLiu,LHan_SA2024_FPan}, from superconductors\cite{CSun_PRB2023_1_JLinder,JAOuassou_PRL2023_JLinder,SBZhang_NC2024_TNeupert,SBanerjee_Arxiv2024_MSScheurer,MMWei_Arxiv2023_JWang} to insulators\cite{JKrempasky_Nature2024_TJungwirth,SLee_PRL2024_CYKim,TOsumi_PRB2024_TSato}. As the vanishing net magnetization in altermagnets avoids effects of long-range magnetic stray fields typical for ferromagnets, this combination opens up promising avenues for applications in magneto-optics\cite{IGray_Arxiv2024_LWu}, spintronics\cite{ZJYJin_Arxiv2023_PYan,HBai_PRL2023_CSong}, and beyond.

Interestingly, symmetry analysis reveals that the electronic structure of altermagnets may in principle provide fertile ground for non-trivial electronic topology, due to the unique interplay between time-reversal and crystalline symmetries\cite{DAntonenko_Arxiv2024_JVenderbos,JXWu_APL2023_XHShao,SDas_Arxiv2024_BRoy,YZhao_Arxiv2024_BYan,RMFernandes_PRB2024_RGPereira}.
Despite extensive experimental and theoretical scrutiny of altermagnets\cite{SLee_PRL2024_CYKim,OFedchenko_SA2024_HJElmers,JKrempasky_Nature2024_TJungwirth,TOsumi_PRB2024_TSato,SReimers_NC2024_MJourdan,ZHLin_Arxiv2024_JZMa,BYChi_PRA2024_XFHan,PRao_Arxiv2024_JKnolle,Yang2024,Zeng2024,Ding2024} and topological semimetals\cite{KManna_NRM2018_CFelser,NPArmitage_RMP2018_AVishwanath,BQLv_RMP2021_HDing,IBelopolski_Science2019_MZHasan,DFLiu_Science2019_YLChen,NMorali_Science2019_HBeidenkopf,CLi_NC2023_OTjernberg,CLi_arxiv2024_2_OTjernberg,NPArmitage_RMP2018_AVishwanath,BQLv_RMP2021_HDing} separately, the simultaneous experimental manifestation of their characteristic phenomena has remained elusive so far. Here we establish CrSb as a topological Weyl semimetal with two distinct types of spin-carrying nodal structures. The giant spin splitting that we find from high resolution angle-resolved photoemission spectroscopy (ARPES) and spin-ARPES measurements aligns well with density functional theory (DFT) calculations on CrSb, as do the bulk Weyl points (WPs) with linear dispersion and ensuing altermagnetic surface Fermi arcs (SFAs) that, as we will show, connect same-spin WPs at the surface.

\begin{figure*}[tbp]
\begin{center}
\includegraphics[width=1\columnwidth,angle=0]{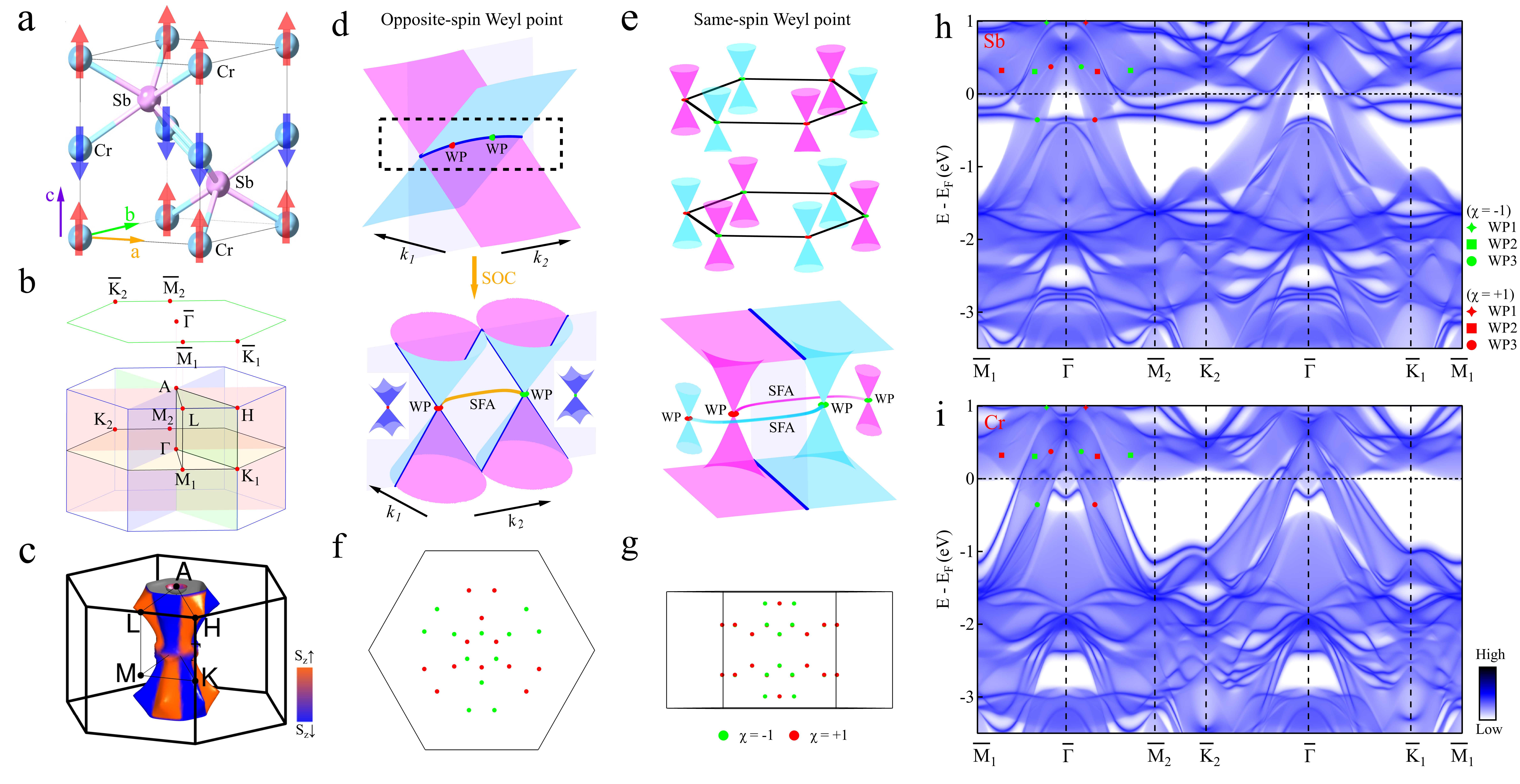}
\end{center}
\caption{\footnotesize\textbf{Crystal structure, WPs and calculated electronic structure of CrSb.} (a) The crystal structure of CrSb with the space group $P6_3/mmc$ (no. 194). (b) The 3D BZ of the original unit cell of CrSb, and the corresponding two-dimensional BZ projected on the (001) plane (green lines). The red, green, blue and orange planes are the four mirror planes in the BZ on which the altermagnetic bands are spin-degenerate in absence of SOC. (c) Calculated bulk Fermi surface in the first BZ. The colormap from blue to orange of Fermi surface sheet represents the expectation value of spin along the $z$ direction. (d-e) Illustration of the two types of WPs: opposite-spin (d) and same-spin (e). Top (f) and side (g) views of the distribution of same-spin WPs in the 3D BZ. Red dots represent nodes with chirality $\chi=+1$, green dots $\chi=-1$; the spin polarizations are indicated by cyan and pink. The blue line represents spin-up and spin-down degeneracy. (h-i) The calculated surface state of (001) on Sb (h) and Cr (i) terminations along $\rm\overline{M}_{1}-\overline{\Gamma}-\overline{M}_{2}-\overline{K}_{2}-\overline{\Gamma}-\overline{K}_{1}-\overline{M}_{1}$. The WPs are marked by the red ($\chi=1$) and green ($\chi=-1$) dots. Star, square and round dots represent the opposite-spin WPs on the high symmetry line/plane (WP1), opposite-spin WPs on general momenta (WP2) and same-spin WPs (WP3), respectively.
}
\label{1}
\end{figure*}

For both its altermagnetic and topological features, the symmetries of CrSb are crucial. Figure~\ref{1}a shows the crystal and magnetic structure of CrSb, which forms a hexagonal structure with the space group $P6_{3}/mmc$ (No.~194)\cite{AISnow_PR1952}. The magnetic space group is $P6'_3/m'm'c$ (No.~194.268) and the spin space group is $P^{-1}{6_{3}/}^{-1}{m}^{1}{m}^{-1}c^{\infty m}1$ with a six-fold screw rotation symmetry connecting Cr-atoms of opposite magnetic moment in real space. The three-dimensional (3D) Fermi surface of CrSb from DFT calculations in Fig.~\ref{1}c illustrates that sixfold rotation connecting opposite-spin sublattices in real space (Fig.~\ref{1}a) also connects opposite-spin electronic states in momentum space (Fig.~\ref{1}c), as is mandatory for altermagnets\cite{LSmejkal_PRX2022_1_TJungwirth,LSmejkal_PRX2022_2_TJungwirth}. The calculated Fermi surface also illustrates the four mirror planes in the Brillouin zone (BZ) on which the altermagnetic bands are spin-degenerate in absence of spin-orbit coupling (SOC), as shown in Fig.~\ref{1}b.    

The altermagnetic spin-splitting allows for topological electronic structures far beyond antiferromagnets with completely spin-degenerate bands\cite{DAntonenko_Arxiv2024_JVenderbos,JXWu_APL2023_XHShao,SDas_Arxiv2024_BRoy,YZhao_Arxiv2024_BYan,RMFernandes_PRB2024_RGPereira}.
We identify two mechanisms for the generation of topological nodes in the electronic structure of CrSb.
The first relies on the observation that spin-splitting of fully spin-polarized altermagnetic bands automatically leads to nodal lines when opposite-spin bands cross. The degenerate states that form the nodal lines have opposite spins. Such crossings in momentum space can be symmetry enforced, for instance at a mirror plane, or be accidental. The effect of SOC in CrSb is to lift this spin-degeneracy for most, but not all momenta on nodal lines. At the remaining Weyl nodes also the spin degeneracy remains, so that the points correspond to doublets with vanishing total spin-projection $S^z=0$ as illustrated in Fig.~\ref{1}d. The altermagnetic symmetries and resulting spin-degeneracies thus provide a very natural setting for such opposite-spin WPs governed by the SOC energy scale. Indeed from DFT we identify 10 groups of opposite-spin WPs within 1 eV from the Fermi energy, see supplementary material for details.

Interestingly, there is also a route that generates purely altermagnetic WPs, reflecting the bulk altermagnetic symmetries. They give rise to same-spin WPs with a total spin-projection of $S^z=\pm 1$ (in units of $\hbar$) and a spin-splitting (energy-splitting at the nodal momentum) governed by the exchange interaction energy scale\cite{LSmejkal_PRX2022_1_TJungwirth,LSmejkal_PRX2022_2_TJungwirth}, as illustrated in Fig.~\ref{1}e. These altermagnetic Weyl nodes thus carry apart from a topological charge also a magnetic quantum number $S^z$, and with it, a finite magnetic moment. It is precisely the altermagnetic crystal symmetry that relates same-spin WPs with different $S^z$. The weak SOC that formally breaks the altermagnetic symmetry in CrSb causes the expectation value $|\langle S^z \rangle |$ to slightly deviate from unity (see supplemental material). These WPs, and the topological surface states they imply, render the topological features of Weyl altermagnets fundamentally distinct from ferro- or antiferromagnetic Weyl semimetals.

For CrSb, where lattice inversion symmetry connects atoms with the same magnetization, the supplementary material provides details of the topological same-spin nodal line without SOC and three groups of same-spin WPs that we identified. Fig.~\ref{1}f-~\ref{1}g shows the distribution of same-spin WPs in momentum space, with their spin-projection quantum number. Fermi-arcs connect the surface projections of these WPs of opposite chirality. As a consequence also their topological surface states are fully spin-polarized in the altermagnetic limit, with additional SOC affecting the details of their dispersion and spectral weight. The electronic structures of both Sb (Fig.~\ref{1}h) and Cr (Fig.~\ref{1}i) terminated surfaces show that the surface states are very extended in momentum space and carry substantial spectral weight. Detailed representations of purely surface related states and the full 3D bulk band structures are provided in the supplementary material. 

\begin{figure*}[tbp]
\begin{center}
\includegraphics[width=1\columnwidth,angle=0]{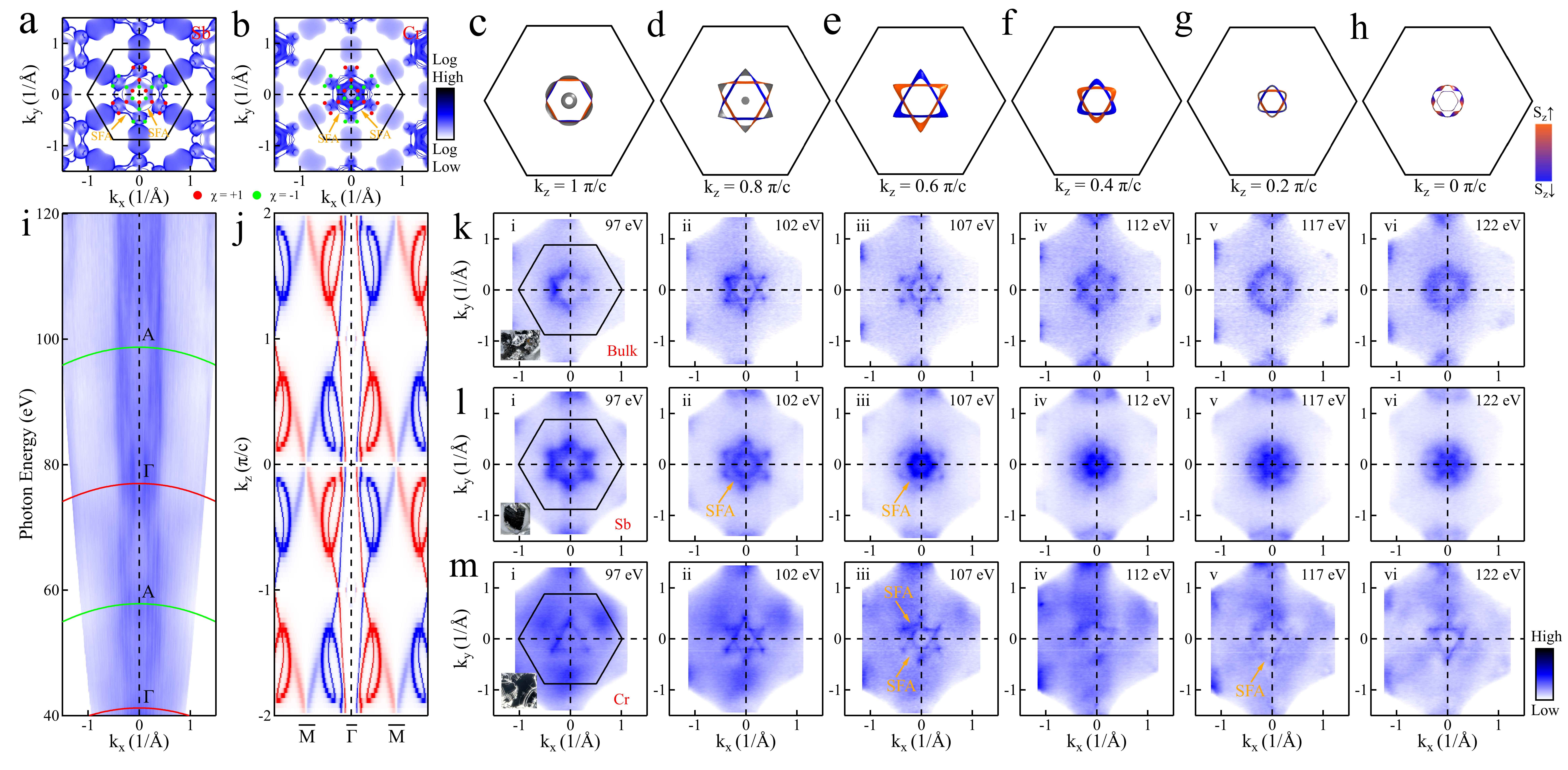}
\end{center}
\caption{\footnotesize\textbf{Fermi surface of CrSb.} (a-b) projected Fermi surfaces for Sb (a) and Cr (b) surface terminations with WPs overlapped. Red dots represent nodes with chirality $\chi=+1$, green dots $\chi=-1$; (c-h) Calculated bulk Fermi surfaces at k$_z$ = 1 $\pi/c$ (c), 0.8 $\pi/c$ (d), 0.6 $\pi/c$ (e), 0.4 $\pi/c$ (f), 0.2 $\pi/c$ (g) and 0 $\pi/c$ (h) planes, integrated over a $\pm0.1$ $\pi/c$ interval. (i) Photon energy dependent ARPES spectral intensity map at Fermi level along $\rm\overline{M}-\overline{\Gamma}-\overline{M}$. (j) corresponding calculated k$_z$-k$_x$ Fermi surface. (k) Photon energy dependent Fermi surfaces measured with photon energies of 97 eV (i), 102 eV (ii), 107 eV (iii), 112 eV (iv), 117 eV (v), 122 eV (vi) on an uneven surface. (l-m) The same measurements on flat areas of Sb (l) and Cr (m) terminated surfaces. The SFAs are marked by orange arrows. The corresponding sample photos are shown in the bottom left corner of k(i), l(i) and m(i).
}
\label{2}
\end{figure*}

To experimentally access the spin-splitting and topological electronic structure of CrSb we use ARPES, which is to a certain extent complicated by CrSb having a 3D electronic structure. In order to establish the correspondence between photon energy and out-of-plane momentum k$_z$, we performed broad-range (40 to 120 eV) photon energy dependent ARPES measurements along the $\rm\overline{M}-\overline{\Gamma}-\overline{M}$ direction. Fig.~\ref{2}i shows the photon energy dependent ARPES spectral intensity map (k$_x$-k$_z$ Fermi surface) at the Fermi level along the $\rm\overline{M}-\overline{\Gamma}-\overline{M}$ direction. From the periodic structure along the k$_z$ direction, the correspondence between the high symmetry points of the BZ along the k$_z$ direction and the photon energy is determined as shown in Fig.~\ref{2}i. To distinguish between bulk and surface states, we measured Fermi surfaces on a relatively uneven sample surface for different k$_z$, corresponding to photon energies from 97 eV to 122 eV (Fig.~\ref{2}k). Since the uneven surface suppresses the contribution of surface states\cite{CLi_NC2023_OTjernberg,CLi_arxiv2024_1_OTjernberg}, these Fermi surfaces mainly derive from the bulk. All their features are captured by the calculated bulk Fermi surface with corresponding k$_z$ (Fig.~\ref{2}c-~\ref{2}h). The Fermi surfaces away from the k$_z$ = 0 $\pi/c$ and 1 $\pi/c$ planes (Fig.~\ref{2}d-~\ref{2}f) have the form of a hexagram consisting of two intersecting equilateral triangular structures. Due to the altermagnetic order in CrSb, the two Fermi surfaces corresponding to the two different triangles, which are very clearly resolved experimentally, have opposite-spin polarizations (Fig.~\ref{2}d-~\ref{2}f). Interestingly, the Sb and Cr terminated surfaces (Fig.~\ref{2}l and ~\ref{2}m) show additional features apart from the bulk Fermi surface, which are marked by orange arrows. As we will detail later, comparison to the calculated surface spectral density (Fig.~\ref{2}a-~\ref{2}b) establishes that their root cause are the topological SFAs that in turn originate from the WPs (see Fig.~S9 in supplementary material for a further comparison of constant energy contours).

\begin{figure*}[tbp]
\begin{center}
\includegraphics[width=1\columnwidth,angle=0]{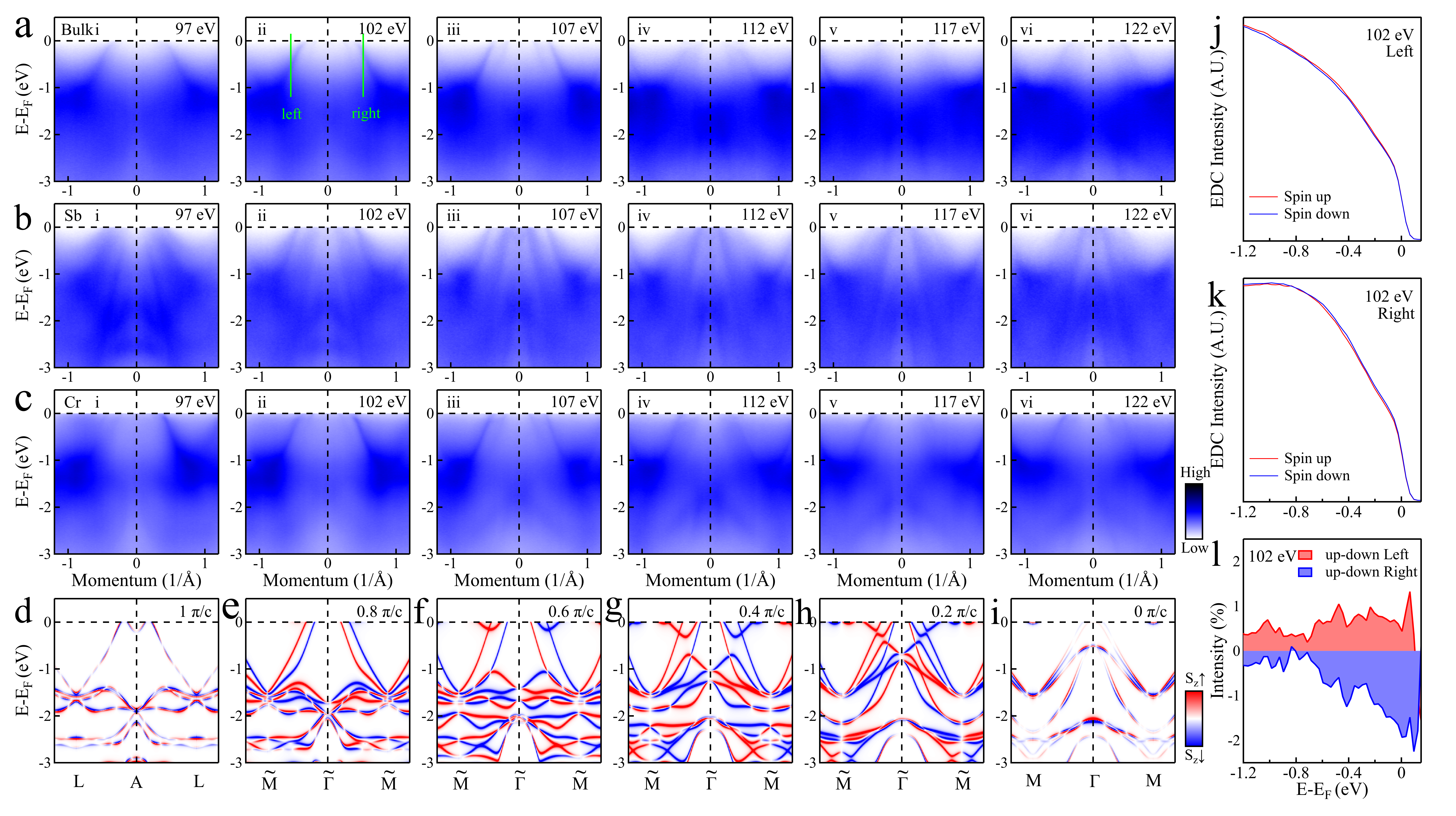}
\end{center}
\caption{\footnotesize\textbf{Band dispersions of CrSb.} (a) The photon energy dependent ARPES spectra along $\rm\overline{M}-\overline{\Gamma}-\overline{M}$ measured with photon energies of 97 eV (i), 102 eV (ii), 107 eV (iii), 112 eV (iv), 117 eV (v) and 122 eV (vi) on an uneven surface. (b-c) The same measurements on flat areas of Sb (b) and Cr (c) terminated surfaces. (d-i) The DFT calculated spin polarized band structures along $\rm\overline{M}-\overline{\Gamma}-\overline{M}$ direction at k$_z$ = 1 $\pi/c$ (d), 0.8 $\pi/c$ (e), 0.6 $\pi/c$ (f), 0.4 $\pi/c$ (g), 0.2 $\pi/c$ (h) and 0 $\pi/c$ (i) planes. The red bands represent a spin-up and the blue bands spin-down along the $z$ direction. (j-k) Spin-polarized EDCs along the left (j) and right (k) cut lines indicated in [a(ii)]. The spin-polarized EDCs were measured with photon energy of 102 eV, while the red (blue) curve corresponds to spin-up (spin-down) along $z$. The EDCs intensity are already normalized in the non-spin polarized region. In order to obtain sufficient spin statistics, more than 30 hours were accumulated in total. (l) Spin-polarized intensity difference between spin-up and spin-down, with the red (blue) colors indicating spin-up (spin-down) polarization for the left (right) bands. 
}
\label{3}
\end{figure*}

First we establish that there is a very large momentum dependent spin-splitting of the bulk bands in CrSb\cite{LSmejkal_PRX2022_1_TJungwirth,Yaqian_Guo_Material_today_physics,SReimers_NC2024_MJourdan,Yang2024,Zeng2024,Ding2024}. 
Fig.~\ref{3}a-~\ref{3}c show the photon energy dependent dispersions along the  $\rm\overline{M}-\overline{\Gamma}-\overline{M}$ direction measured with photon energies from 97 eV to 122 eV with the interval of 5 eV for an uneven sample surface (Fig.~\ref{3}a) and flat Sb (Fig.~\ref{3}b) and Cr (Fig.~\ref{3}c) terminated surfaces. The corresponding k$_z$ dependent calculated spin-polarized band structures are shown alongside it (Fig.~\ref{3}d-~\ref{3}i). These bands are spin degenerate on the k$_z$ = 0 $\pi/c$ and 1 $\pi/c$ planes. Due to the altermagnetism in CrSb the bands between these two k$_z$ planes (Fig.~\ref{3}e-~\ref{3}h) exhibit spin splitting. The calculated bands and their spin splittings are in very good agreement with the ARPES measurements [Fig.~\ref{3}a-~\ref{3}c, (ii-v)]. We determine the maximum energy band splitting to be up to 1 eV (see Fig.~S10 in the supplementary material). To further verify that the split bands are spin polarized, we performed spin resolved ARPES measurements with a photon energy of 102 eV, as shown in Fig.~\ref{3}j-~\ref{3}l. Fig.~\ref{3}j and ~\ref{3}k are the spin-polarized energy distribution curves (EDCs) along the left and right cut lines in Fig.~\ref{3}a(ii). The spin-polarized intensity difference between spin-up and spin-down of left and right bands is shown in Fig.~\ref{3}l. Due to the existence of a large non-spin polarized background, originating from a Shirley background, the measured spin polarization observed in the EDCs is limited. Despite this, a difference in energy-dependent spin polarization intensity between spin-up and spin-down of left and right bands (Fig.~\ref{3}l) can still be observed. This indicates that the left and right energy bands [Fig.~\ref{3}a(ii)] have different spin projections along the $z$ direction, in agreement with theory.

\begin{figure*}[tbp]
\begin{center}
\includegraphics[width=1\columnwidth,angle=0]{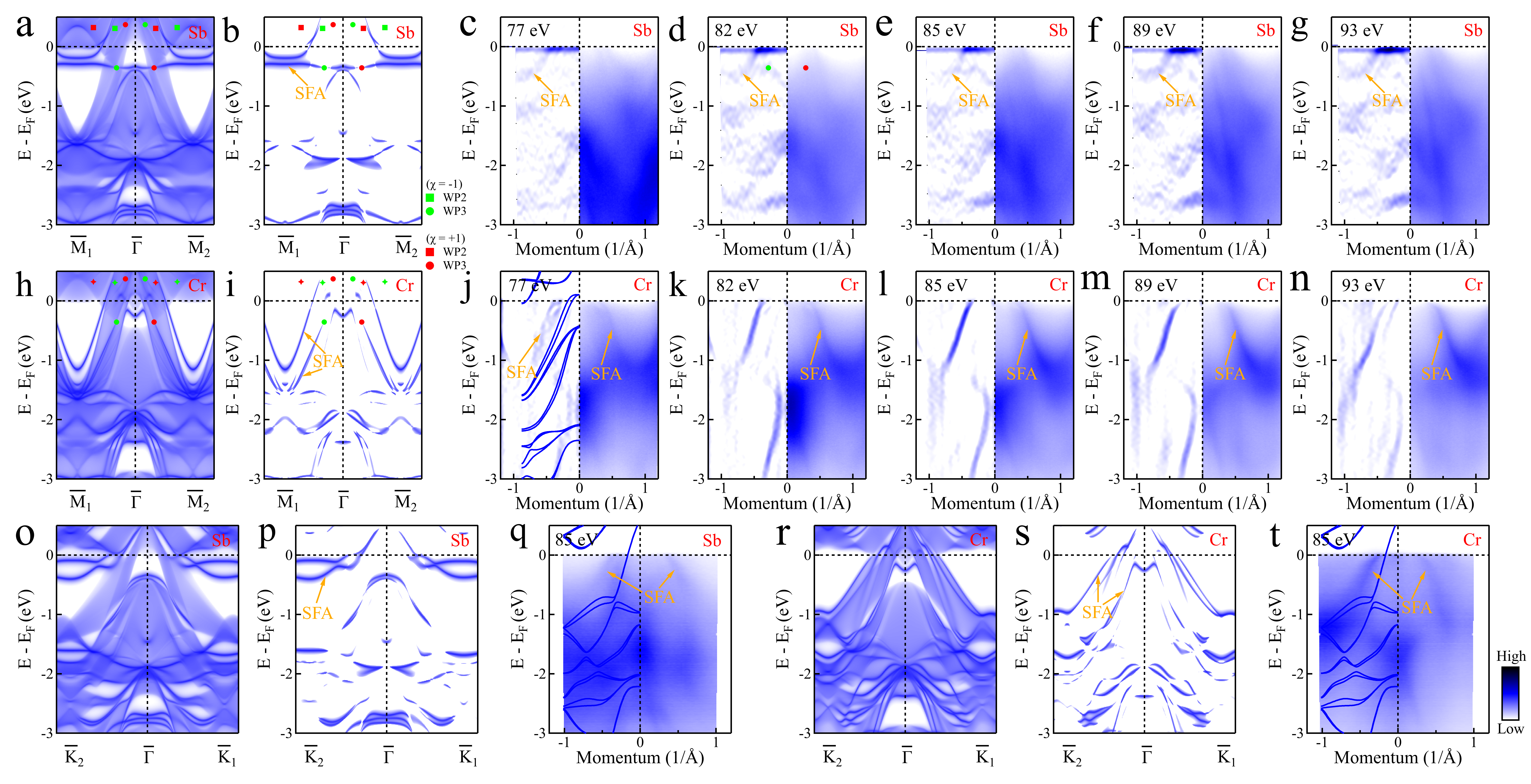}
\end{center}
\caption{\footnotesize\textbf{Weyl fermions and Fermi arcs in CrSb.} (a) Calculated total spectral density along $\rm\overline{M}-\overline{\Gamma}-\overline{M}$ for CrSb terminated by a Sb (001) surface. (b) surface electronic structure as in (a) but with bulk spectral weight subtracted. The WPs together with the chirality are marked by red ($\chi=1$) and green ($\chi=-1$) dots. Square and round dots represent the opposite-spin WPs at general momenta (WP2) and same-spin WPs (WP3), respectively. (c-g) The photon energy dependent spectra on a Sb terminated surface with photon energies of 77 eV (a), 82 eV (b), 85 eV (c), 89 eV (d), 93 eV (e) under LH polarization along $\rm\overline{M}-\overline{\Gamma}-\overline{M}$. (h-n) The same calculations and measurements as (a-g) for the Cr terminal surface. In (j) on the left also the bulk bands for k$_z$ = 0 $\pi/c$ are shown. (o) Calculated total spectral density along $\rm\overline{K}-\overline{\Gamma}-\overline{K}$ of the Sb surface. (p) The same but with bulk contribution subtracted. (q) Measured band dispersion along the $\rm\overline{K}-\overline{\Gamma}-\overline{K}$ direction of a Sb terminated surface with photon energy of 85 eV. (r-t) The same calculations and measurements as (o-q) but on a Cr terminated surface. The bulk calculations for k$_z$ = 0.4 $\pi/c$ are also shown on the left side of (q) and (t). The SFAs are marked by orange arrows. The left sides in (c-g) correspond to the EDC second derivative and (j-n) show the MDC second derivative, enhancing the visibility of the bands.
}
\label{4}
\end{figure*}

Having identified and quantified the altermagnetic spin-splitting in CrSb, we focus on the resulting topological properties. Both ARPES measurements and DFT calculations show that CrSb is metallic with a complex Fermi surface. In an energy range around from $-1$ eV to 1 eV around the Fermi level, we identify 13 groups of WPs in the calculated bands, as listed in Table.~1 of the supplementary material. All WPs in CrSb arise from time-reversal symmetry breaking and their relative positions are given by the symmetries in the magnetic space group $P6_3'/m'm'c$ (No.~194.268) reflecting their unique altermagnetic nature. Many WPs are along or near the $\rm\overline{\Gamma}-\overline{M}$ direction (see Fig.~S2 in supplementary material) -- this is the high-symmetry line whose little co-group allows non-zero chirality. The pair of WPs in the $\rm\overline{\Gamma}-\overline{M}$ direction below the Fermi level (marked by round dots in Fig.~\ref{4}a-~\ref{4}b and ~\ref{4}h-~\ref{4}i) are same-spin WPs. Long SFAs connect neighboring WPs of opposite chirality $\chi = \pm 1$ and identical spin-projection across the BZ. This topological surface state happens to appear in a region without bulk states so that it may be clearly discerned in ARPES. Even if the presence of these SFAs is dictated by topology, the shape of the surface spectra for two different terminations differ due to the lack of inversion and mirror symmetry on the surface, as shown in Fig.~\ref{4}. Due to the large distance between two WPs projected on the 001 surface, the arc is even detectable along $\rm\overline{\Gamma}-\overline{K}$ as shown in Fig.~\ref{4}o-~\ref{4}p. 

To track these WPs and SFAs experimentally, we conducted photon energy dependent ARPES measurements along $\rm\overline{M}_{1}-\overline{\Gamma}-\overline{M}_{2}$ on Sb (Fig.~\ref{4}c-~\ref{4}g) and Cr (Fig.~\ref{4}j-~\ref{4}n) terminated surfaces. The band features measured on Sb (marked by orange arrows in Fig.~\ref{4}c-~\ref{4}g) and Cr (marked by orange arrows in Fig.~\ref{4}j-~\ref{4}n) surfaces exhibit negligible photon energy dependence [momentum distribution curve (MDC) analysis of band features measured on Cr terminated surface see Fig.~S11 in supplementary material], evidencing that it corresponds to a surface state. Comparison with the corresponding calculated surface states, shows that their band features (marked by orange arrows in Fig.~\ref{4}c-~\ref{4}g and Fig.~\ref{4}j-~\ref{4}n) agree well with the SFAs (marked by orange arrows in Fig.~\ref{4}b and ~\ref{4}i), indicating that the SFAs are intrinsic and robust. As pointed out earlier, the SFAs are also observed in the 2D momentum maps at the Fermi energy (Fig.~\ref{2}l and ~\ref{2}m). Furthermore, the DFT calculations show the k$_z$ location of WPs at an energy of -0.357 eV (round dots in Fig.~\ref{4}a-~\ref{4}b and ~\ref{4}h-~\ref{4}i) is $\pm$0.27 $\pi/c$. This corresponds to a photon energy of about 82 eV. 
We show the WP in the band dispersion panel measured with 82 eV (Fig.~\ref{4}d) and indeed it is located  right at the intersection of the bands. In addition, we also measured the band dispersion along $\rm\overline{\Gamma}-\overline{K}$  with photon energy of 85 eV on Sb (Fig.~\ref{4}q) and Cr (Fig.~\ref{4}t) surfaces (for more detailed photon energy dependent measurements along this direction see Fig.~S12 in supplementary material). Also along these high-symmetry lines, SFAs (denoted by orange arrows in Fig.~\ref{4}q and ~\ref{4}t) are observed for both surface terminations, reflecting the large portion of the surface BZ traversed by the SFA, consistent with the calculations. 

Thus we have not only quantified the altermagnetic spin-spliting of bands in CrSb with high-resolution and spin-resolved ARPES measurements, but also established it as a topological Weyl semi-metal, with concomitant altermagnetic topological surface states. Whereas in general Weyl nodes carry only a topological charge, corresponding to their chirality $\chi$, in altermagnets they also carry spin-projection $S^z$. Besides opposite-spin nodes with zero spin-projection, altermagnetic CrSb also has same-spin nodes with spin-projection $\pm \hbar$ and chirality-spin locking. For each set of $n$ symmetry-related altermagnetic Weyl nodes one can define a locking invariant $\zeta = \sum_{i=1,n} \chi_i S^z_i$. For CrSb $\zeta=0$ by symmetry, but interestingly other magnetic symmetry groups allow sets of altermagnetic nodes with non-zero integer locking invariants. The distribution of same-spin nodes in 3D momentum space reflects all altermagnetic symmetries as do their spin-polarized topological surface states, the altermagnetic Fermi arcs. Spin-orbit coupling weakly breaks altermagnetic symmetries and consequently the quantization of the spin-projection. As these properties derive from the interplay of symmetry and topology, they are generic for Weyl altermagnets. These findings may well imply that the unique bulk altermagnetic spin-transport properties are promoted by topology to also become properties of the Fermi arcs at the surface, which can render same-spin Weyl altermagnets interesting interface materials for spintronics. These insights not only spotlight the distinctive altermagnetic attributes of CrSb but also its potential to induce novel physics and applications in the area of topological materials.\\

\noindent {\bf Methods}\\
\noindent{\bf Sample} The CrSb single crystals were grown by the chemical vapor transport (CVT) method. A stoichiometric ratio of chromium and antimony powders, together with iodine of 2.5 mg/ml as the transport agent, were mixed and sealed in an evacuated quartz ampoule. The ampoule was slowly heated and finally exposed to a temperature gradient of 925°C to 900°C where the CVT preceded for one week, then naturally cooled down to room temperature. CrSb crystals in size of ~5 mm with regular shapes and shiny surfaces were obtained. 

\noindent{\bf ARPES Measurements} High-resolution ARPES measurements were performed at the Bloch beamline of MAX IV and at the I05 beamline of the Diamond synchrotron light source. The total energy resolution (analyzer and beamline) was set at 15$\sim$20 meV for the measurements. The angular resolution of the analyser was $\sim$0.1 degree. The beamline spot size on the sample was about 10 $\mu$m$\times$12 $\mu$m at the Bloch beamline of MAX IV and about 70 $\mu$m$\times$70 $\mu$m at the I05 beamline of the Diamond synchrotron. The samples were cleaved {\it in situ} and measured at about 18 K at the Bloch beamline of MAX IV and about 10 K at the I05 beamline of the Diamond synchrotron in ultrahigh vacuum with a base pressure better than 1.0$\times$10$^{-10}$ mbar. The spin-resolved ARPES (SARPES) measurements were performed at the Bloch B-branch beamline of MAX IV with photon energy of 102 eV. The samples were cleaved {\it in situ} and measured at about 77 K at the Bloch B-batch beamline of MAX IV. The total energy resolution (analyzer and beamline) was set at $\sim$50 meV for the measurements. The angular resolution of the analyser was $\sim$1.3 degree.

\noindent{\bf DFT calculations} 
Full relativistic spin-polarized electronic structure and Fermi surface calculations are  done using the full potential local orbital code FPLO~\cite{KKoepernik_PRB1999_HEschrig,KKoepernik_PRB2023_JVDBrink}, on a $k$-mesh of $20\times20\times13$ points. We use the generalized gradient approximation to the exchange and correlation potential by Perdew, Burke, and Ernzerhof (PBE).
The surface state and WPs calculations are performed based on the DFT calculation from VASP \cite{GKresse_PRB1996_JFurthmuller} employing the projector augmented wave method \cite{PEBlochl_PRB1994}. The Brillouin zone is sampled with a $10\times10\times8$, with Gamma-centered $k$-point. The energy cutoff of the plane wave basis is set to 550 eV. The Hubbard term was introduced and set to be 0.8 eV in the $d$ orbitals of Cr-atom in the DFT framework (DFT+U) in order to account for the electron-electron correlation. The Wannier based Hamiltonian is symmetrized based on the maximally localized Wannier functions generated by the WANNIER90 interface \cite{NMarzai_PRB1997_DVanderbilt}. The projectors are $d$ orbitals of Cr and $p$ orbitals of Sb atoms with the well-fitted region from -2 to 2 eV. To locate the WPs and calculate the chirality, Wanniertools is implemented \cite{QSWu_CPC2018_AASoluyanov}.

\noindent {\bf Data Availability}

\noindent The authors declare that all data supporting the findings of this study are available within the paper and its Supplementary Information files.

\vspace{3mm}

\noindent {\bf Acknowledgements}\\
The work presented here was financially supported by the Swedish Research council (2019-00701) and the Knut and Alice Wallenberg foundation (2018.0104). 
We thank Ulrike Nitzsche for technical assistance. We acknowledge financial support by the Deutsche Forschungsgemeinschaft (DFG, German Research Foundation), through SFB 1143 (Project ID 247310070), project A05, Project No. 465000489, and the W\"urzburg-Dresden Cluster of Excellence on Complexity and Topology in Quantum Matter, ct.qmat (EXC 2147, Project ID 390858490).
M.G.V. thanks support from the Deutsche Forschungsgemeinschaft (DFG, German Research Foundation) GA3314/1-1 -FOR 5249 (QUAST) and  to the Spanish Ministerio de Ciencia e Innovacion grant PID2022-142008NB-I0.
M.L.H. thanks the support from the Alexander von Humboldt Foundation and the useful discussion and help of Iñigo Robredo. 
Z.L.L thanks the support from the Youth Innovation Promotion Association of Chinese Academy of Sciences (No. 2021008).
H.L. was supported by the Swedish Research Council (VR, grant 2018-00313), the Wallenberg Academy Fellows program of the Knut and Alice Wallenberg Foundation (2018.0460) and the G\"oran Gustafsson Foundation for Research in Natural Sciences and Medicine.
We acknowledge MAX IV Laboratory for time on Beamline BLOCH under Proposal 20230262 and 20231119. Research conducted at MAX IV, a Swedish national user facility, is supported by the Swedish Research council under contract 2018-07152, the Swedish Governmental Agency for Innovation Systems under contract 2018-04969, and Formas under contract 2019-02496.
\vspace{3mm}

\noindent {\bf Author Contributions}\\
C.L. and H.L. proposed the project. C.L., M.L.H., H.L., O.T. and J.V.D.B. conceived the project. C.L. carried out the ARPES experiments with the assistance from Y.W. and W.Y.C.. M.L.H. and O.J. contributed to the band structure calculations. Z.L.L.  contributed to CrSb crystal growth. C.L. contributed to software development for data analysis and analyzed the data. C.L., M.L.H., H.L., O.T. and J.V.D.B. wrote the paper. B.T., M.L., C.P. and T.K. provided the beamline support. C.L, M.L.H., Y.W., H.L., C.F., M.V., O.J., O.T. and J.V.D.B. contributed to the scientific discussions. All authors participated in and commented on the paper.

\noindent {\bf Competing Interests}\\
The authors declare no competing interests.

\newpage

\end{document}


\centerline{\large{Supplementary Material for}}

\title{Topological Weyl Altermagnetism in CrSb}

\author{Cong Li$^{1,*,\sharp}$, Mengli Hu$^{2,*}$, Zhilin Li$^{3,*}$, Yang Wang$^{1}$, Wanyu Chen$^{1}$, Balasubramanian Thiagarajan$^{4}$, Mats Leandersson$^{4}$, Craig Polley$^{4}$, Timur Kim$^{5}$, Hui Liu$^{6,\sharp}$, Cosma Fulga$^{2,7}$, Maia G. Vergniory$^{8,9}$, Oleg Janson$^{2}$, Oscar Tjernberg$^{1,\sharp}$, Jeroen van den Brink$^{2,7,\sharp}$
}

\affiliation{
\\$^{1}$Department of Applied Physics, KTH Royal Institute of Technology, Stockholm 11419, Sweden
\\$^{2}$Leibniz Institute for Solid State and Materials Research, IFW Dresden, Helmholtzstraße 20, 01069 Dresden, Germany
\\$^{3}$Beijing National Laboratory for Condensed Matter Physics, Institute of Physics, Chinese Academy of Sciences, Beijing 100190, China
\\$^{4}$MAX IV Laboratory, Lund University, 22100 Lund, Sweden
\\$^{5}$Diamond Light Source, Harwell Campus, Didcot, OX11 0DE, United Kingdom
\\$^{6}$Department of Physics, Stockholm University, AlbaNova University Center, 10691 Stockholm, Sweden
\\$^{7}$Würzburg-Dresden Cluster of Excellence Ct.qmat, Technische Universitat Dresden, 01062, Dresden, Germany
\\$^{8}$Max Planck Institute for Chemical Physics of Solids, 01187 Dresden, Germany
\\$^{9}$Donostia International Physics Center, 20018 Donostia–San Sebastian, Spain
\\$^{*}$These authors contributed equally to the present work.
\\$^{\sharp}$Corresponding authors: conli@kth.se, hui.liu@fysik.su.se, oscar@kth.se, j.van.den.brink@ifw-dresden.de
}

\pacs{}

\maketitle

\tableofcontents
\newpage

\section{Projected Wannier function}

\begin{figure*}[h]
\begin{center}
\includegraphics[width=0.8\columnwidth,angle=0]{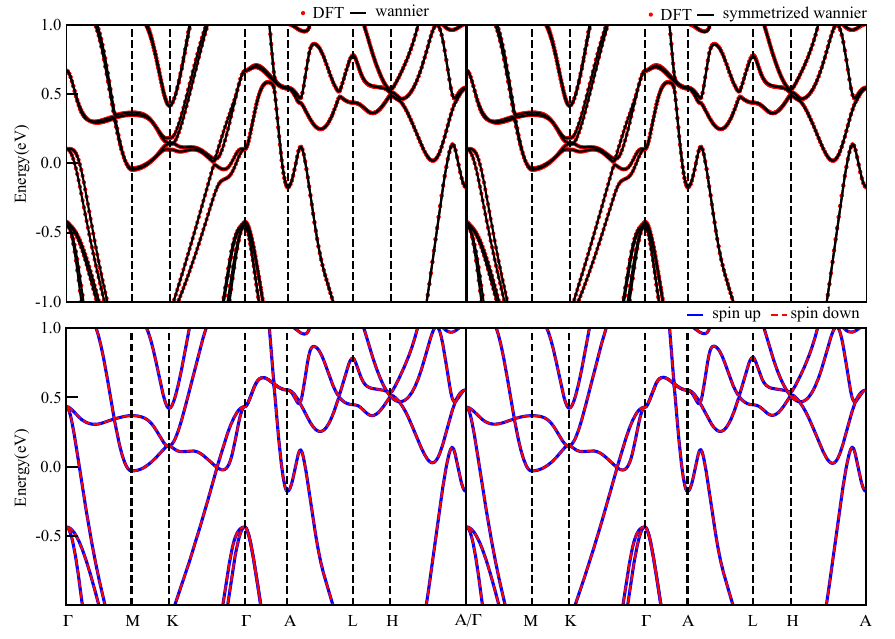}
\end{center}
\caption{\label{S1}\textbf{Wannier function projection comparison with DFT calculation} Upper panel: Band structure calculated by DFT labelled by red solid dots. Fitted Wannier functions and the symmetrized Wannier functions are presented on the left and right panels denoted by black solid lines, respectively. Lower panel: Band structure calculated by DFT without SOC. The left panel presents the band structure calculated based on symmetrized Wannier functions, and the right panel is the DFT calculation results. The calculated bands are along $\Gamma-M-K-\Gamma-A-L-H-A$.
}
\end{figure*}

To study the topological properties of CrSb, we projected the VASP-calculated bands onto a Wannier basis using the maximally-localized Wannier functions formalism implemented in Wannier90. Fig.~\ref{S1} shows that the Fourier-transformed Wannier Hamiltonian agrees very well with the DFT band structure. For further analysis, we additionally symmetrized the resulting hopping parameters such that they respect all symmetries of the magnetic space group $P6_3'/m'm'c$ (No. 194.268). We note that this symmetrization does not give rise to any discernible changes in the spectrum. All further discussions of the topology of CrSb are based on results obtained for this symmetrized Hamiltonian.

We also performed the DFT calculation without SOC and did the corresponding Wannier projection. As shown in the lower panel in Fig.~\ref{S1}, the symmetrized Wannier Hamiltonian also shows good consistency with the DFT results. To arrive at a symmetrized Hamiltonian, we averaged all symmetrically equivalent hopping parameters as explained in Ref.~\onlinecite{Symmetrized_TB}.
\newpage

\section{WPs locations}

In our full relativistic calculations, i.e.\ with the spin-orbit coupling included, we have searched the Weyl points (WPs) spanning from the energy range of 1 eV around the Fermi level. In this way, we found 13 groups of WPs, for which we evaluated the corresponding spin expectation values. 
The respective spin operator in orbital and spin basis reads: $\sigma_z=I_N\bigotimes \sigma_z$, where $N$ is the number of total number of basis divided by 2. The total spin expectation values for WPs take the form: $\expval{S_z} = \sum_i^2 \mel{\phi_i}{\hat{\sigma_z}}{\phi_i}/2$. 
We identify the type of WPs as same-spin when $|\expval{S_z}|> 0.9$ and opposite-spin when $|\expval{S_z}|\leq 0.2$.
In this way, we classify them according to their spin values in the fully spin polarized (without SOC) limit: 1) opposite-spin WPs enforced by spin space group symmetry, 2) accidental opposite-spin WPs, and 3) same-spin WPs. For the first two types, WPs are labeled as Opposite-1 and Opposite-2 in the following tables. All 13 groups of WPs are sorted by their magnetic little co-groups.\\

\begin{figure*}[h]
\begin{center}
\includegraphics[width=0.85\columnwidth,angle=0]{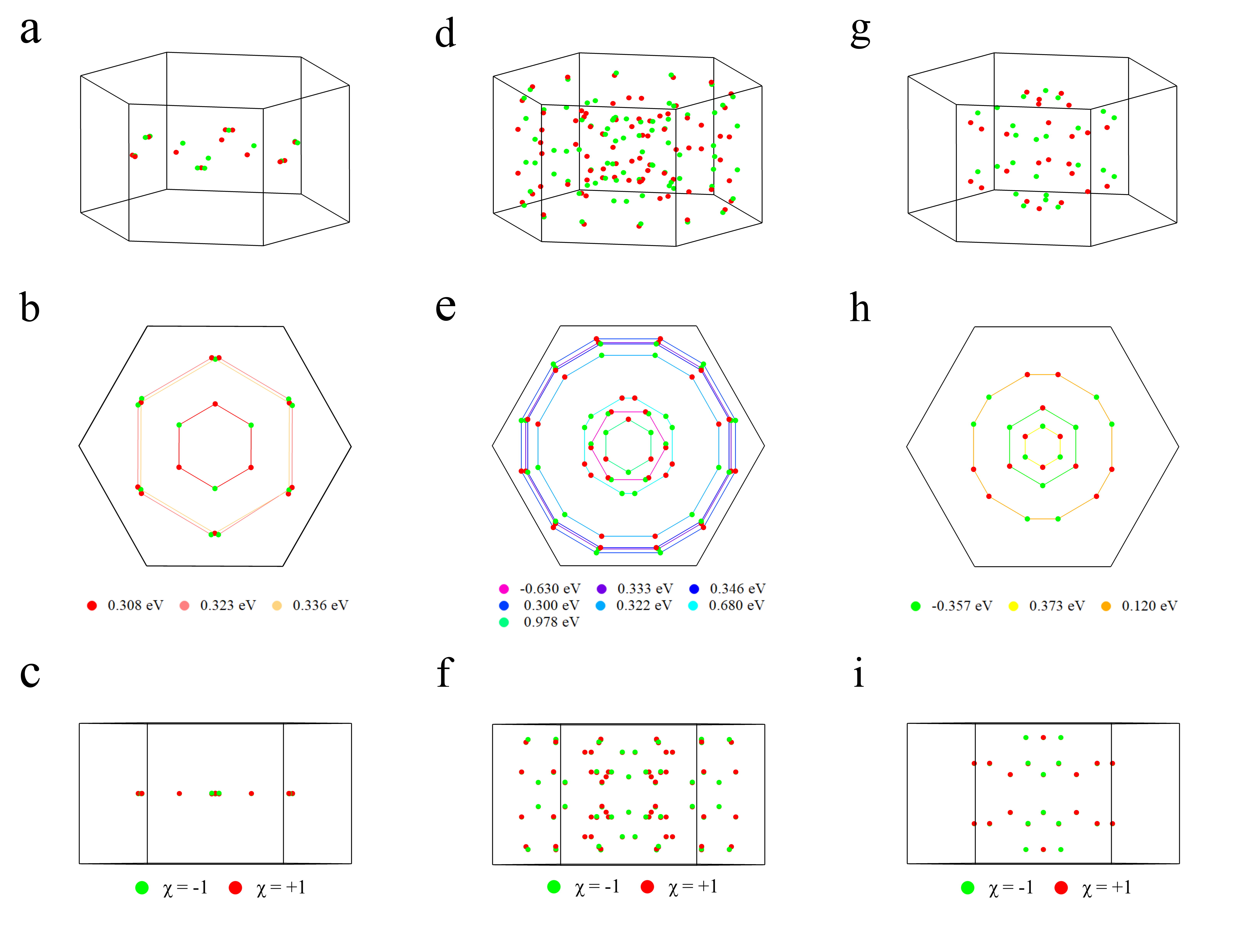}
\end{center}
\caption{\label{S2}\textbf{WPs of CrSb.} (a) The distribution of opposite spin WPs on the high symmetry line/plane in the 3D BZ, with the red dots representing points with chirality $\chi$ = +1 and green dots representing $\chi$ = -1. (b-c) The top (b) and side (c) views of (a). The same group of WPs is connected with colored lines. Different colors represent the different energy positions of WPs. (d-f) The similar plot as (a-c) but corresponding to the opposite spin WPs on general momentum. (g-i) Similar plots as (d-f) but corresponding to the same spin WPs.
}
\end{figure*}

\begin{table}[H]
\centering
\begin{tabular}{c c c c c c} 
 \hline
 Number of WPs & location (u,v,w) & Energy & Chirality & Type & $\expval{S_z}$\\ [0.5ex] 
 \hline\hline
 2.1 & (-0.18323,   0.09917,  -0.16014) & -0.62992 & -1 & Opposite-2 & -0.174 \\ 
 3.1 & (-0.26333,  -0.07510,   0.21453) & \ 0.12050 & 1  & Same-spin & \ 0.994\\
 2.2 & (0.36769,  -0.52436,  -0.16000) & \ 0.30000 & 1 & Opposite-2 & \ 0.007\\ 
 2.3 & (0.14573,  -0.50290 ,  0.37165) & \ 0.33314 & -1 & Opposite-2 & -0.020\\ 
 2.4 & (0.13637 ,  0.35652,   0.39145) & \ 0.34602 & -1 & Opposite-2& \ 0.036\\
 2.5 & (-0.44259 ,  0.13094,  -0.08619) & \ 0.32175 & -1 & Opposite-2 & \ 0.002 \\
 2.6 & (-0.21457,   0.03116,  -0.30092) & \ 0.68000 & 1 & Opposite-2 & -0.038\\ 
 \hline
\end{tabular}
\caption{WPs belonging to little co-group: $1$. There are 24 symmetry connected WPs that share the same relation of location and chirality where WPs at (u,v,w),(v,-u-v,w),(-u-v,u,w),(-v,-u,-w),(-u,u+v,-w),(u+v,-v,-w),(u,v,-w),(v,-u-v,-w),(-u-v,u,-w),(-v,-u,w), (-u,u+v,w),(u+v,-v,w) share the chirality and at (-u,-v,-w),(-v,u+v,-w),(u+v,-u,-w),(v,u,w),(u,-u-v,w),(-u-v,v,w), (-u,-v,w),(-v,u+v,w),(u+v,-u,w),(v,u,-w),(u,-u-v,-w),(-u-v,v,-w) host opposite chirality.}
\label{table:1}
\end{table}
\begin{table}[H]
\centering
\begin{tabular}{c c c c c c} 
 \hline
 Number of WPs & location & Energy & Chirality & Type & $\expval{S_z}$ \\ [0.5ex] 
 \hline\hline
 3.2 & (0.16200,   0.00000,  -0.13500) & -0.35700 & -1 & Same-spin & \ 0.981\\ 
 3.3 & (0.08556,   0.00000,   0.39873) & \ 0.37304 & 1 & Same-spin & -0.952\\ 
 2.7 & (-0.11043,   0.00000,  -0.12601) & \ 0.97800 & 1 & Opposite-2 & -0.017\\
 1.1 & (-0.37754,   0.01763,   0.00000) & \ 0.33580 & 1 & Opposite-1 & 0.000\\ 
 \hline
\end{tabular}
\caption{WPs are belong to little co-group: $2'$. There are 12 symmetry connected WPs and for No. 3.2, No. 3.3, and No. 2.7, the symmetry connected WPs locations are (u,0,w),(0,-u,w),(-u,u,w),(0,-u,-w),(-u,u,-w),(u,0,-w),(-u,0,-w) have same chirality, and (0,u,-w), (u,-u,-w),(0,u,w),(u,-u,w),(-u,0,w) host opposite chirality to the first half WPs. No. 1.1 WPs are in the $k_z=0$ plane and their Weyl pairs are (u,v,0),(v,-u-v,0),(-u-v,u,0),(-u,-v,0),(-v,u+v,0),(u+v,-u,0),(-v,-u,0),(-u,u+v,0),(u+v,-v,0),(v,u,0),(u,-u-v,0),(-u-v,v,0) where the first half share the same chirality and the remaining half is opposite.}
\label{table:1}
\end{table}
\begin{table}[H]
\centering
\begin{tabular}{c c c c c c} 
 \hline
 Number of WPs & location & Energy & Chirality & Type  & $\expval{S_z}$\\ [0.5ex] 
 \hline\hline
 1.2 & (-0.17637   0.00000   0.00000) &\ 0.30800 & 1 & Opposite-1 & -0.017\\ 
 1.3 & (0.00000   0.36304   0.00000) & \ 0.32330 & -1 & Opposite-1 & \ 0.008\\ 
 \hline
\end{tabular}
\caption{WPs belonging to little co-group: $2'2'2$. There are 6 symmetry connected WPs share the same relation of location and chirality where WPs are at (u,0,0),(-u,0,0),(0,u,0),(0,-u,0),(-u,u,0),(u,-u,0) and first half of them share the same chirality and the remaining half have opposite chirality of the first.}
\label{table:1}
\end{table}
To visualize all the WPs, we divided these WPs into three categories and plotted their positions in the three-dimensional (3D) Brillouin zone in Fig.~\ref{S2}. Fig.~\ref{S2}a shows the 3D distribution of opposite spin WPs on the high symmetry line/plane in the 3D BZ (red dots representing points with chirality $\chi$ = +1 and green dots representing $\chi$ = -1). Fig.~\ref{S2}b-\ref{S2}c show the top view and side view of Fig.~\ref{S2}a which gives a more complete view of the distribution of WPs. The same group of WPs is connected with colored lines. Different colors represent the different energy positions of WPs. Fig.~\ref{S2}d-\ref{S2}f and Fig.~\ref{S2}g-\ref{S2}i show similar plots of the WPs as Fig.~\ref{S2}a-\ref{S2}c but corresponding to opposite spin WPs for general momenta (Fig.~\ref{S2}d-\ref{S2}f) and same spin WPs (Fig.~\ref{S2}g-\ref{S2}i). \\

\newpage

\section{Calculation of CrSb surface states along high-symmetry lines}

\begin{figure*}[h]
\begin{center}
\includegraphics[width=0.85\columnwidth,angle=0]{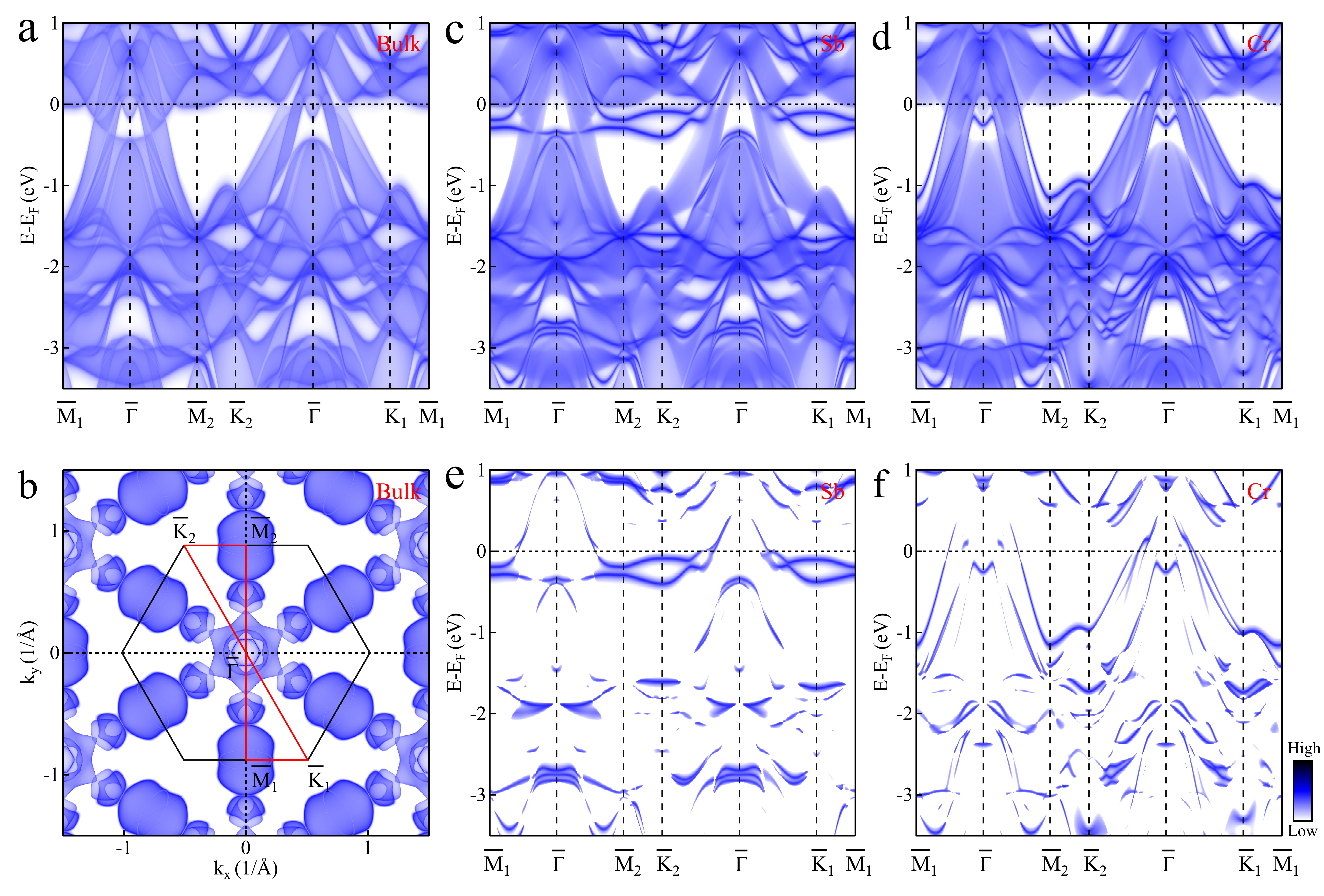}
\end{center}
\caption{\label{S3}\textbf{Surface state calculation along high symmetry lines} (a), (c), (d) are the calculated bulk and surface spectral densities along $\rm\overline{M}_1-\overline{\Gamma}-\overline{M}_2-\overline{K}_2-\overline{\Gamma}-\overline{K}_1-\overline{M}_1$. The high symmetry points are defined in (b). (b) is the bulk density of state in the BZ at the Fermi level. (e-f) are the surface spectral densities subtracted by bulk densities. 
}
\end{figure*}

Fig.~\ref{S3}a-\ref{S3}d show the spectral density from semi-infinite CrSb calculation. Using the iterative Green's function technique, the bulk (Fig.~\ref{S3}a-\ref{S3}b), Sb (Fig.~\ref{S3}c), and Cr (Fig.~\ref{S3}d) terminated surface spectrum are obtained. To further analyze the surface states, Fig.~\ref{S3}e and Fig.~\ref{S3}f are the results with bulk weight subtracted.
\newpage

\section{k$_z$ dependent band structures of CrSb}

\begin{figure*}[h]
\begin{center}
\includegraphics[width=1\columnwidth,angle=0]{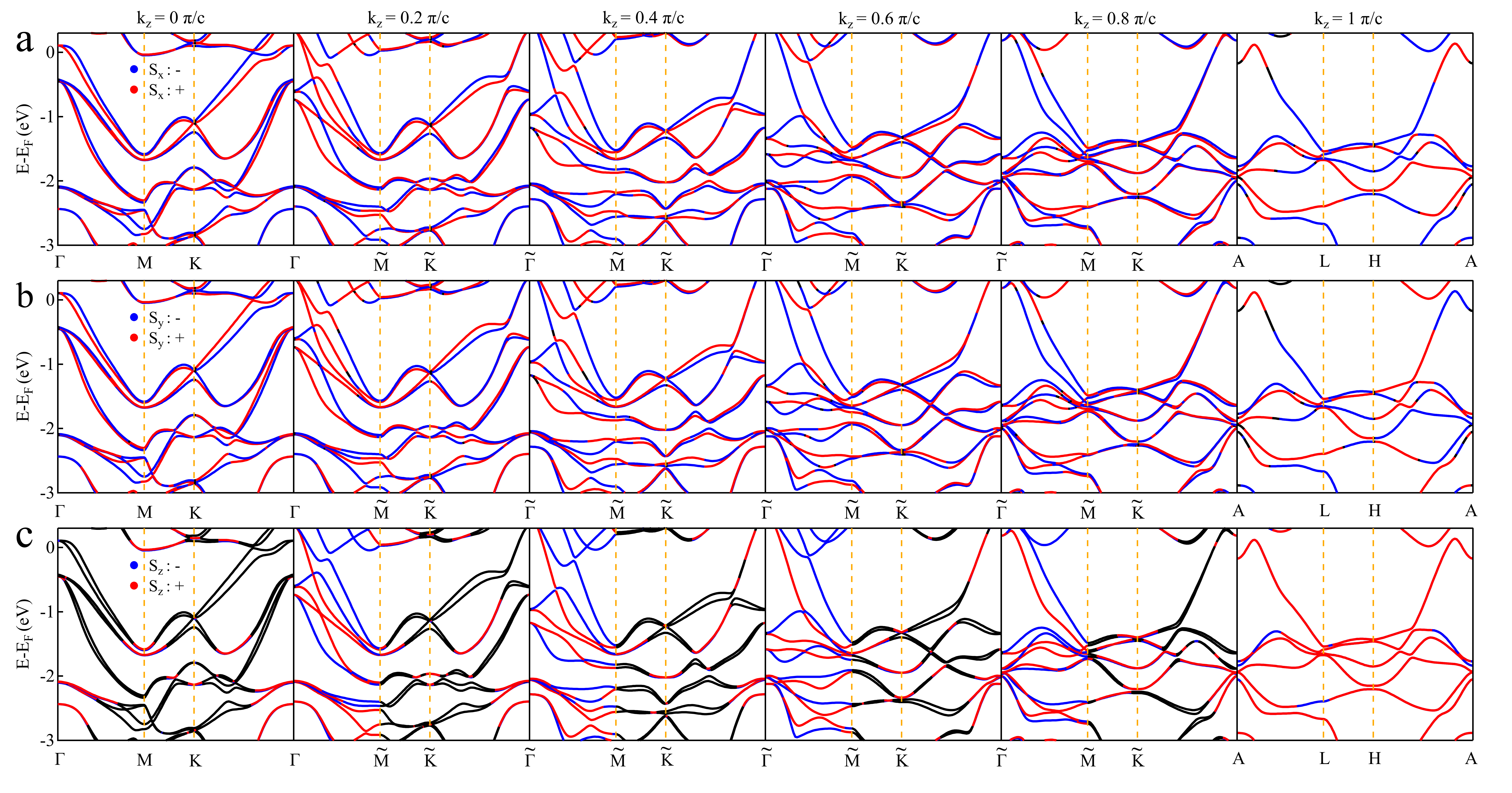}
\end{center}
\caption{\label{S4}\textbf{K$_z$ dependent spin polarized band structures of CrSb.} (a-c) k$_z$ dependent spin polarized band structures along $\rm\widetilde{\Gamma}-\widetilde{M}-\widetilde{K}-\widetilde{\Gamma}$ with spin projected along the $x$ (a), $y$ (b) and $z$ (c) directions. The red (blue) bands represent  spin-up (spin-down) along the corresponding directions. Black bands represent spin degenerate bands.
}
\end{figure*}

Fig.~\ref{S4} shows the in-plane band dispersion for various k$_z$, with the respective spin projections onto the $x$ (Fig.~\ref{S4}a), $y$ (Fig.~\ref{S4}b) and $z$ (Fig.~\ref{S4}c) axes. As discussed in the main text, in the absence of SOC, doubly degenerate bands are confined to the four mirror planes. With SOC, degeneracies in the three vertical mirror planes are lifted, while those in the k$_z$ = 0 plane remain. Nonsymmorphic symmetries~\cite{Nonsymm_3} of the magnetic space group, e.g.\ a two-fold screw axis combined with time reversal, give rise to  k$_z$-dependent degeneracy, which also happens in transition metal dichalcogenides~\cite{Nonsymm_1,Nonsymm_2}. Along the $\widetilde{K}-\widetilde{\Gamma}$ high symmetry line, nonzero spin projection along the $z$ direction is not allowed due to the mirror symmetry. Also WPs are not allowed on the mirror plane in CrSb. The band structure analysis for different k$_z$ also reveals that SOC plays a role in breaking the nodal lines/surfaces in CrSb.\\

\newpage

\section{Surface states of CrSb}

\begin{figure*}[h]
\begin{center}
\includegraphics[width=.95\columnwidth,angle=0]{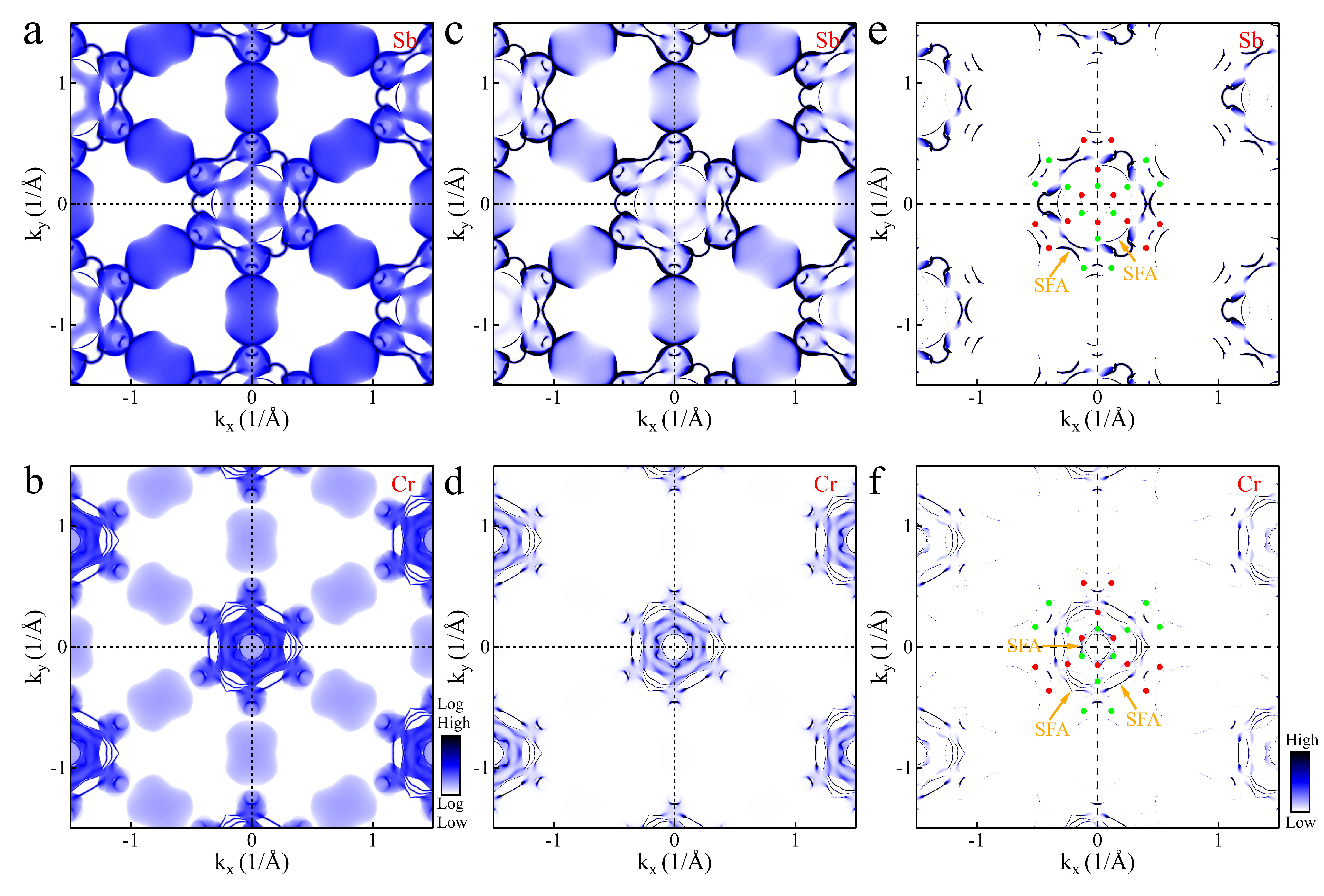}
\end{center}
\caption{\label{S5}\textbf{Calculated surface spectral densities for CrSb with log and linear color scales.} (a), (c) and (b), (d) are Sb and Cr terminated surfaces,respectively, displayed with log (a-b) and linear (c-d) color scales. (e-f) The DFT calculated surface projected Fermi surface on Sb (e) and Cr (f) terminated surfaces after subtraction of the bulk states. The orange arrows mark the SFAs.
}
\end{figure*}

Fig.~\ref{S5} shows the calculated surface spectral density at the Fermi level of the Sb- (Fig.~\ref{S5}a and \ref{S5}c) and Cr-terminated (Fig.~\ref{S5}b and \ref{S5}d)  surfaces. We use a logarithmic (linear) color scale in Fig.~\ref{S5}a and \ref{S5}b (Fig.~\ref{S5}c and \ref{S5}d). These figures are also shown in the main-text as Fig.~2a and 2b. To emphasize contributions of the surface states, bulk states have been subtracted following the equation: $\rho_s(\epsilon,\bf{k})-\rho_b(\epsilon,\bf{k})$ from Fig.~\ref{S5}c and \ref{S5}d and the pure surface density results is shown in Fig.~\ref{S5}e and \ref{S5}f. The SFAs are marked by orange arrows. For each projected WP on the (001) surface, there are two WPs with identical chirality at the same position in the 2D BZ, that give rise to two SFAs. SFAs labelled in Fig.~\ref{S5}e and \ref{S5}f originate from WPs No. 3.2 and No. 3.3, whose positions along $\overline{\Gamma}-\overline{M}$ and denoted. \\

\newpage

\section{Experimental determination of the surface termination in CrSb}

\begin{figure*}[h]
\begin{center}
\includegraphics[width=0.85\columnwidth,angle=0]{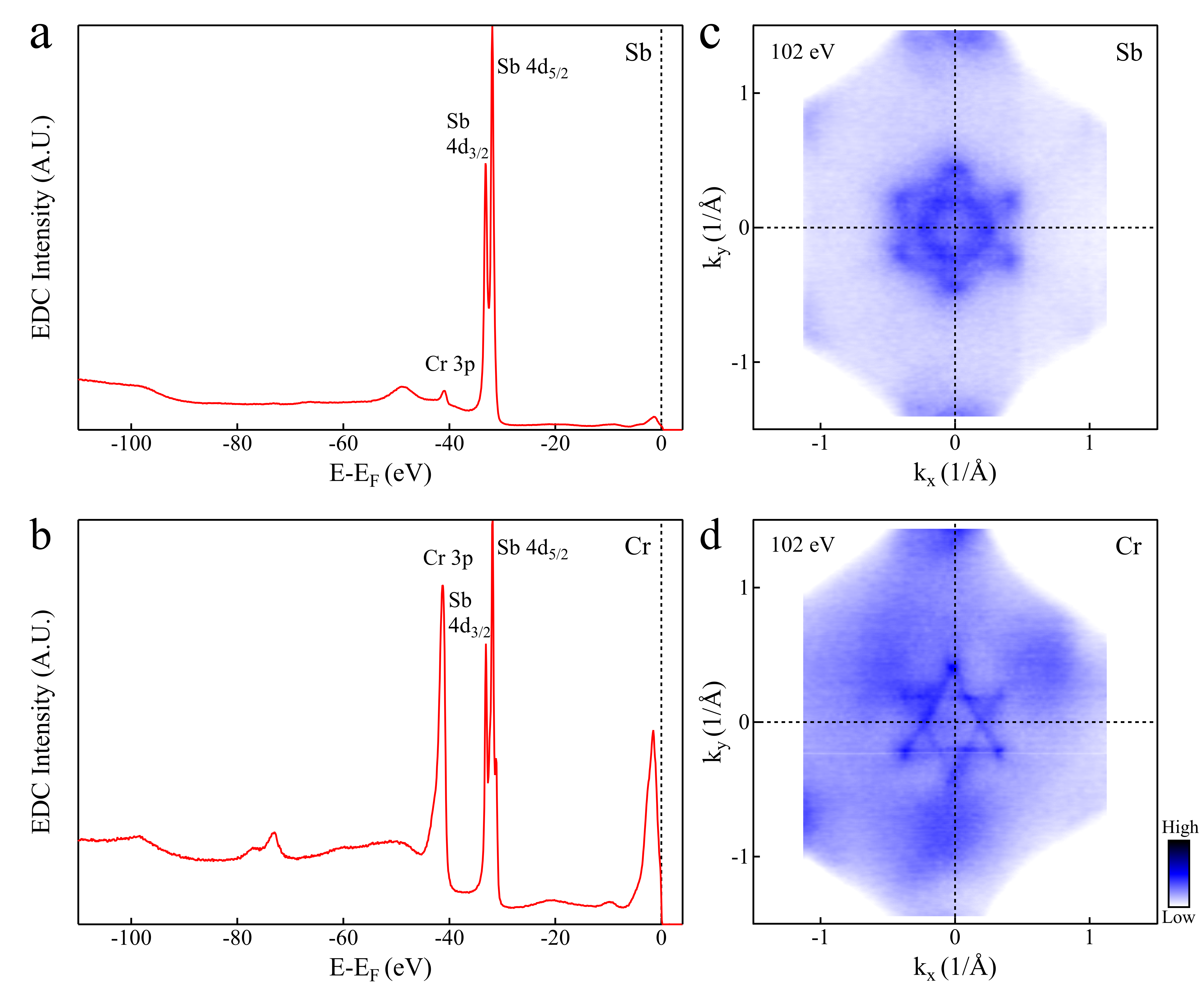}
\end{center}
\caption{\label{S6}\textbf{Determination of the surface terminations of CrSb.} (a-b) XPS measurements on the (0001) surface measured on Sb (a) and Cr (b) terminated surfaces. The peak intensity of the Cr 3p state for Sb termination is much lower than for Cr termination. (c-d) The Fermi surfaces measured for terminations corresponding to the XPS measurements in (a-b).
}
\end{figure*}

The (0001) surfaces of CrSb can terminate with a triangular-lattice monolayer of either Cr or Sb atoms. 
Experimentally, we observed two dissimilar Fermi surface structures, as shown in Fig.~\ref{S6}c and \ref{S6}d. It is natural to assume that these two Fermi surfaces correspond to different surface terminations. To verify this conjecture, we conducted X-ray photoelectron core-level spectroscopy (XPS) measurements (Fig.~\ref{S6}a-\ref{S6}b) on the two terminations corresponding to the measured Fermi surfaces (Fig.~\ref{S6}c-\ref{S6}d). The result is that the peak intensity of Cr $3p$ states in Fig.~\ref{S6}a is much lower than that in Fig.~\ref{S6}b, indicating that the XPS measurement in Fig.~\ref{S6}c (Fig.~\ref{S6}d) corresponds to the Sb (Cr) terminated surface.

\begin{figure*}[h]
\begin{center}
\includegraphics[width=1\columnwidth,angle=0]{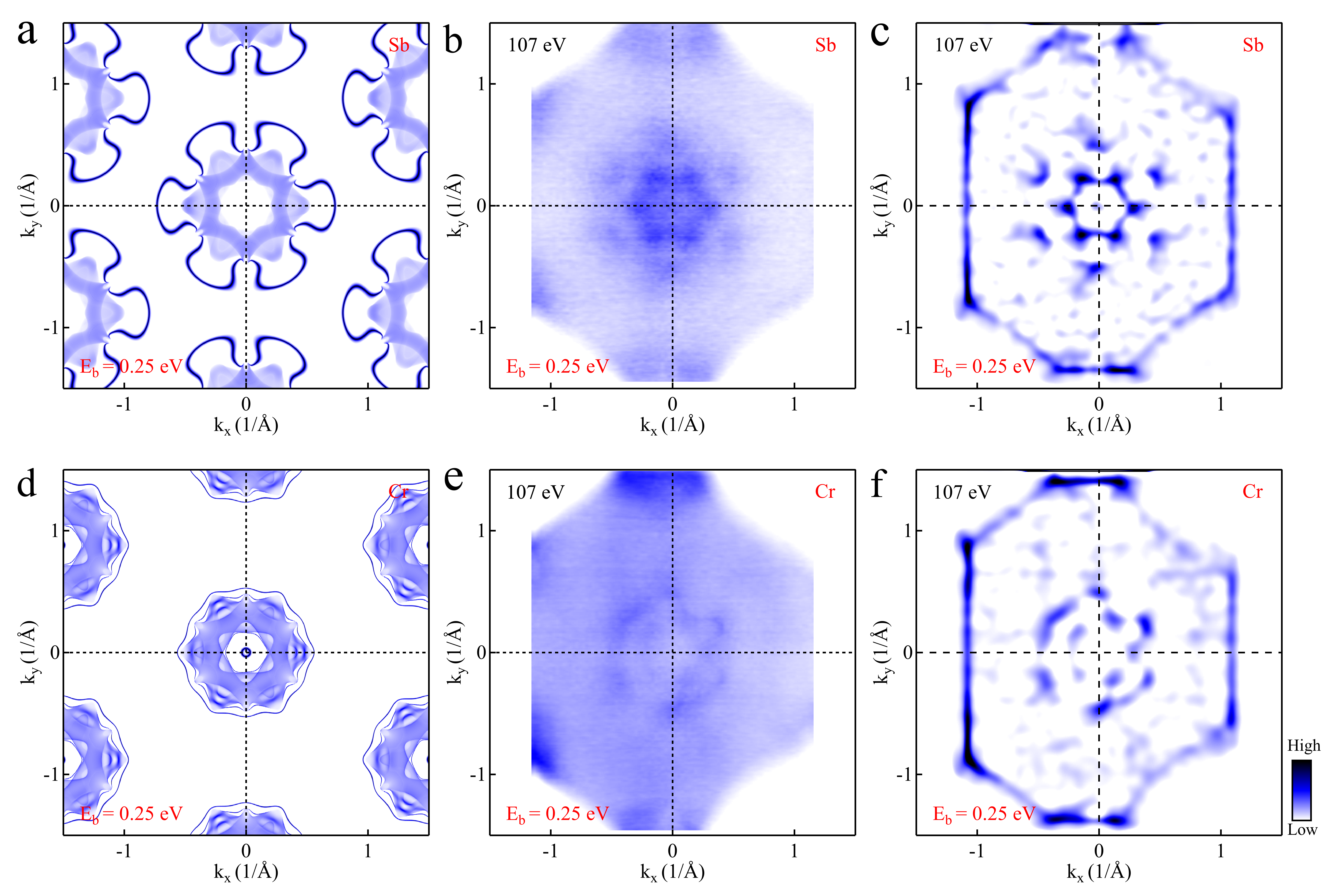}
\end{center}
\caption{\label{S07}\textbf{Comparison of surface density of states at E$_b$ = 0.25 eV.} (a) and (d): Calculated surface density of states at E$_b$ = 0.25 eV on the Sb terminated and Cr terminated surfaces. (b) and (e) are measured Fermi surfaces at E$_b$ = 0.25 eV on the Sb terminated (b) and Cr terminated (e) surfaces with a photon energy of 107 eV. (c) and (f) are the two dimensional (2D) curvature image of (b) and (e), respectively.}
\end{figure*}

After establishing the correspondence between the surface termination and the measured Fermi surfaces, we further compared the latter (Fig.~\ref{S07}b-\ref{S07}c and Fig.~\ref{S07}e-\ref{S07}f) with the calculated surface spectral density at the binding energy of 0.25 eV  (Fig.~\ref{S07}a-\ref{S07}b). The calculated spectra of two different terminations are indeed dissimilar, and are both in good agreement with the experiment. For the Sb-terminated surface, the calculated spectral density shows distinct petal-like SFA whose traces are also visible in the curvature of the experimental intensities.

\newpage

\section{Distinguishing the surface state of the Cr terminated surface}

\begin{figure*}[h]
\begin{center}
\includegraphics[width=1\columnwidth,angle=0]{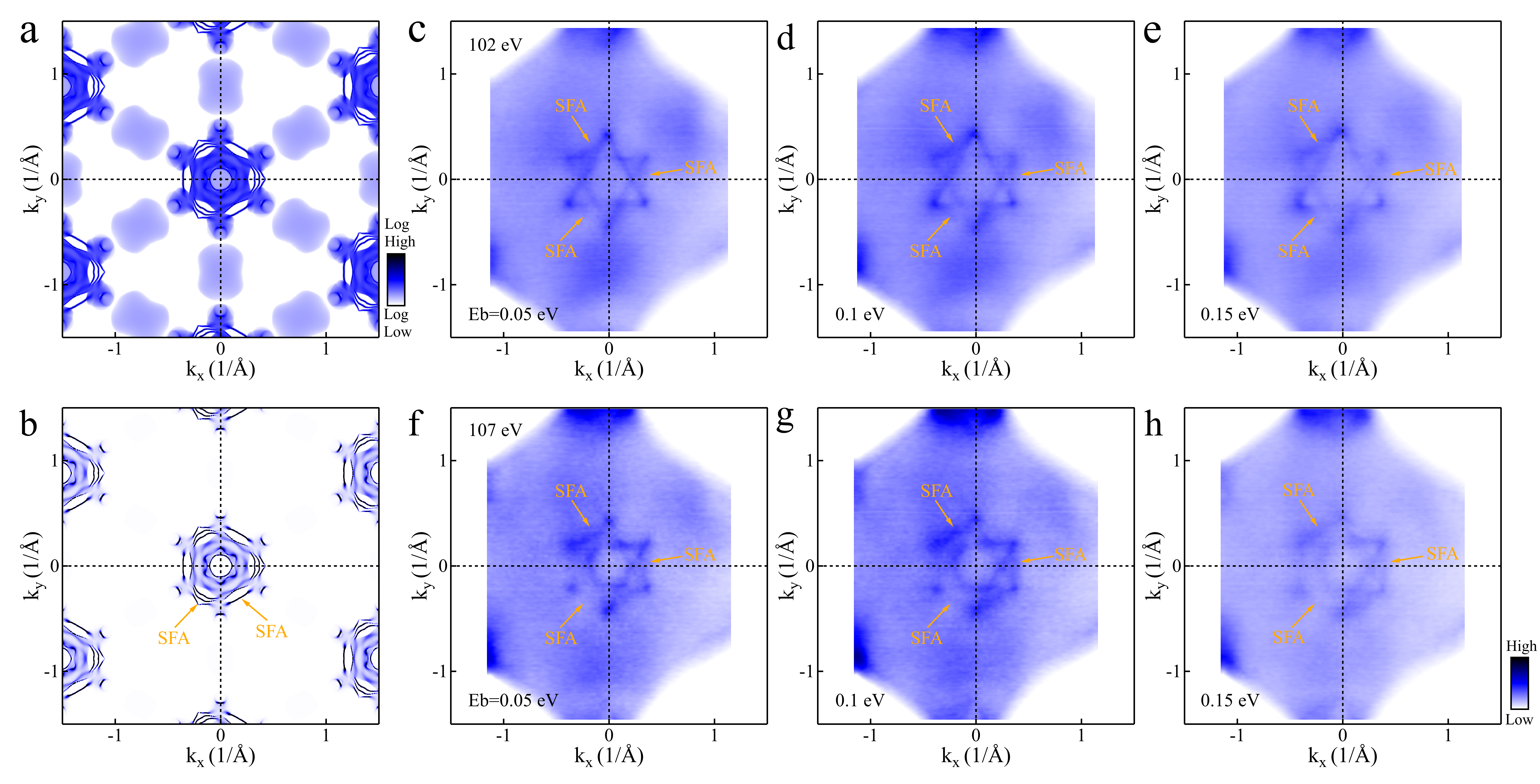}
\end{center}
\caption{\label{S8}\textbf{Distinguishing the surface states of the Cr terminated surface.} (a-b) Surface projected DFT calculated Fermi surfaces of the Cr terminated surface displayed with log (a) and linear (b) color scale. (c-e) The constant energy contours for the Cr terminated surface measured with a photon energy of 102 eV at the binding energy of 0.05 eV (c), 0.1 eV (d) and 0.15 eV (e). (f-h) The constant energy contours for the Cr terminated surface measured with a photon energy of 107 eV at the binding energy of 0.05 eV (f), 0.1 eV (g) and 0.15 eV (h). The surface states are marked by orange arrows.
}
\end{figure*}

Fig.~\ref{S8}a and \ref{S8}b show the calculated surface state at the Fermi level of the Cr terminated surface displayed with a log (Fig.~\ref{S8}a) and a linear (Fig.~\ref{S8}b) color scale, respectively. The SFAs of the Cr terminated surface are marked by orange arrows in Fig.~\ref{S8}b. Comparing these measurements, similar features can be observed in the measured constant energy contours, marked by orange arrows in Fig.~\ref{S8}c-\ref{S8}h, indicating the SFA nature of them.\\

\newpage

\section{Estimation of the energy scale of the maximum band spin splitting in CrSb}

\begin{figure*}[h]
\begin{center}
\includegraphics[width=0.75\columnwidth,angle=0]{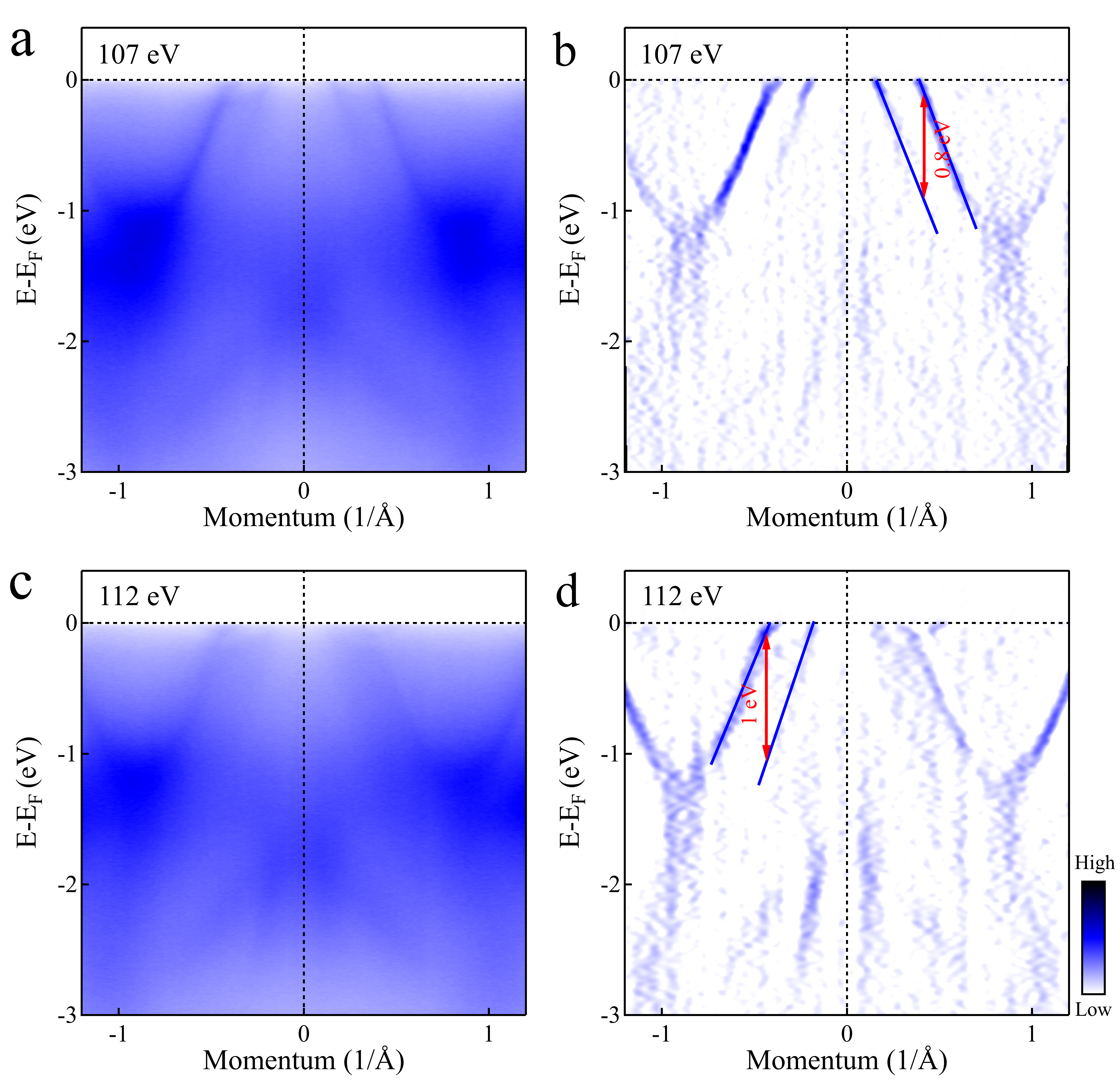}
\end{center}
\caption{\label{S9}\textbf{Estimating the energy scale of the band spin splitting.} (a) Band dispersion measured with a photon energy of 107 eV along $\rm\overline{M}-\overline{\Gamma}-\overline{M}$. (b) The second derivative image of (a). (c-d) Similar measurements as (a-b) but with a photon energy of 112 eV. The measured band spin splitting is 0.8 eV for 107 eV and 1 eV for 112 eV photon energy, respectively.
}
\end{figure*}

According to the DFT bulk band calculations, the maximum band spin splitting is found around the k$_z$ = 0.4 $\pi/c$ plane. Fig.~\ref{S9} shows the band dispersion measured with photon energy of 107 eV (see Fig.~\ref{S9}a-\ref{S9}b) and 112 eV (see Fig.~\ref{S9}c-\ref{S9}d), corresponding to the k$_z$ planes around 0.4 $\pi/c$. Fig.~\ref{S9}b and \ref{S9}d are the second derivative images of Fig.~\ref{S9}a and \ref{S9}c, enhancing the band contrast. The so-estimated band spin splitting amounts to about 0.8 eV for 107 eV (Fig.~\ref{S9}b) and 1 eV for 112 eV (Fig.~\ref{S9}d).\\

\newpage

\section{Observation of two domain structures in CrSb}

\begin{figure*}[h]
\begin{center}
\includegraphics[width=1\columnwidth,angle=0]{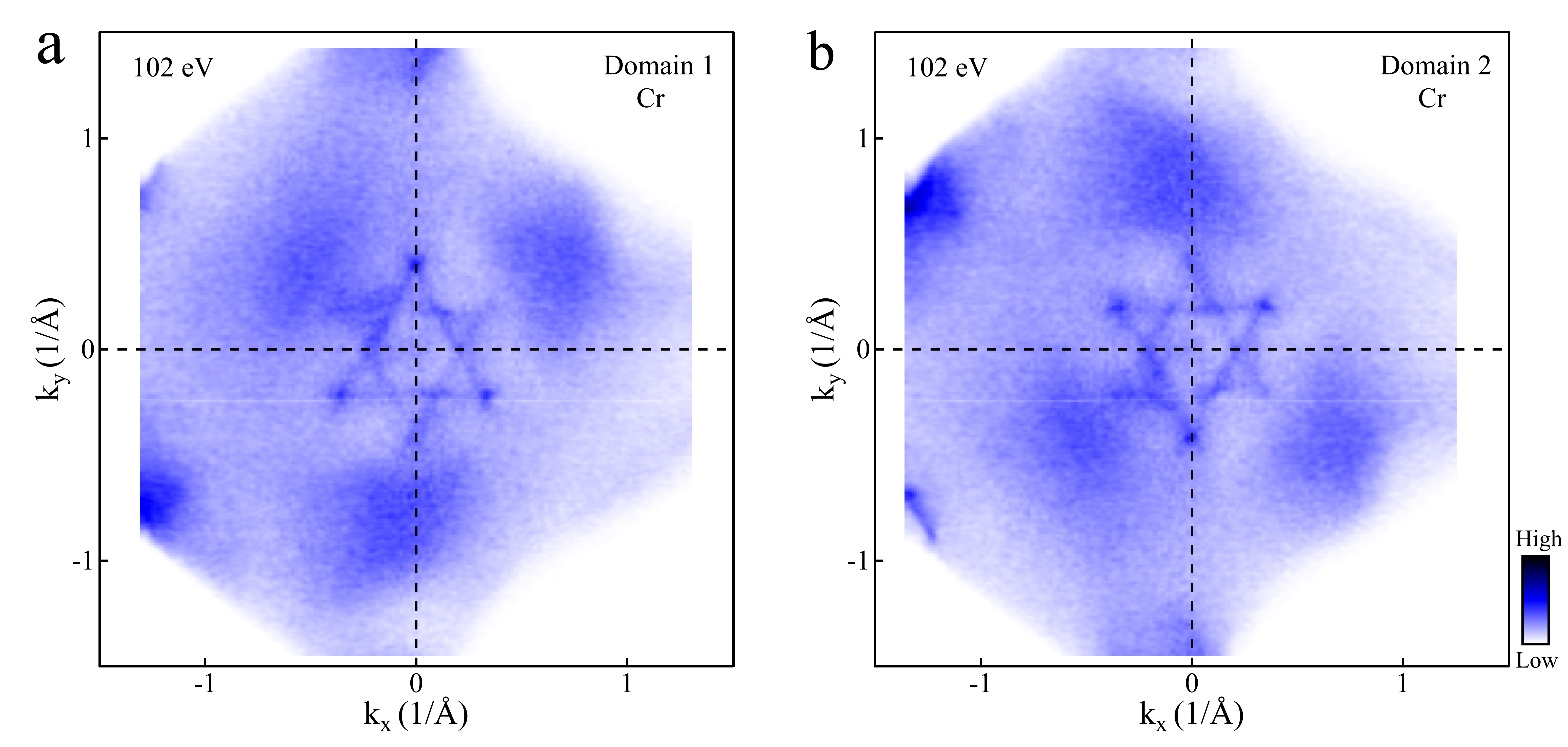}
\end{center}
\caption{\label{S7}\textbf{Two domain structures on the Cr terminated surface of CrSb.} (a-b) Fermi surface measured with photon energy of 102 eV on domain 1 (a) and domain 2 (b) of the Cr terminated surface.
}
\end{figure*}

When performing spatial scans on the Cr terminated surface, we observe that the measured Fermi surface flips at certain positions on the sample surface (Fig.~\ref{S7}a-\ref{S7}b). This indicates that there are two domain structures on the Cr terminated surface of CrSb.\\

\newpage

\section{Photon energy dependent band dispersion measurements}

\begin{figure*}[h]
\begin{center}
\includegraphics[width=0.85\columnwidth,angle=0]{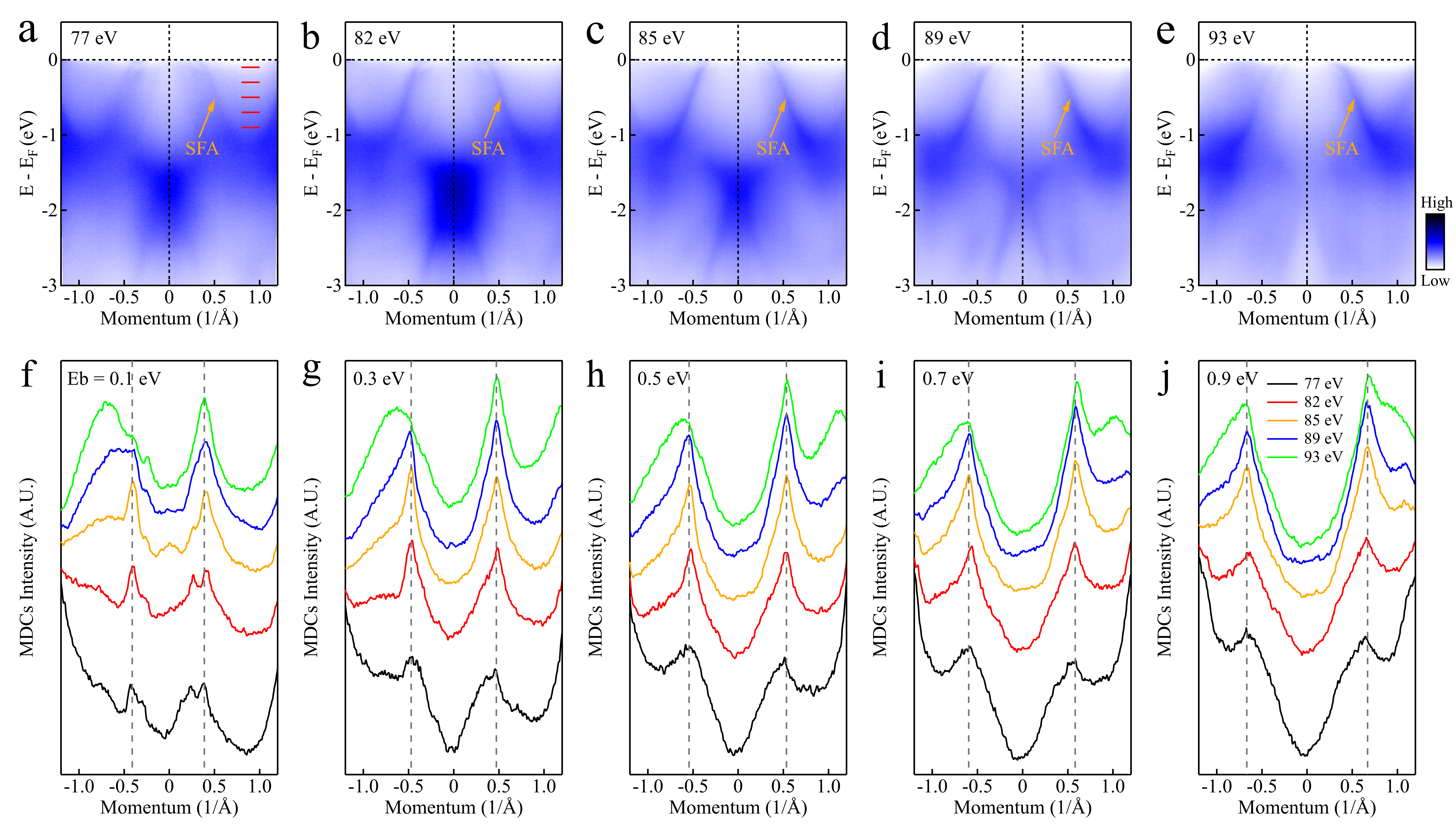}
\end{center}
\caption{\label{S10}\textbf{Photon energy dependent band dispersion along the $\rm\overline{\Gamma}-\overline{M}$ direction on the Cr terminated surface.} (a-e) Photon energy dependent band dispersion measurements along $\rm\overline{\Gamma}-\overline{M}$ on the Cr terminated surface with photon energies of 77 eV (a), 82 eV (b), 85 eV (c), 89 eV (d) and 93 eV (e). (f-j) The photon energy dependent MDCs extracted from (a-e) at the binding energies of 0.1 eV (f), 0.3 eV (g), 0.5 eV (h), 0.7 eV (i) and 0.9 eV (j). The MDC peak positions are marked by dashed vertical lines in (f-j). Energies of the MDCs are marked in (a) by red lines.
}
\end{figure*}

Fig.~\ref{S10} shows the photon energy dependent band dispersion measurements along the $\rm\overline{\Gamma}-\overline{M}$ direction on the Cr terminated surface. 
In order to determine the surface state properties of the band features marked by orange arrows in Fig.~\ref{S10}a-\ref{S10}e, we analyzed the momentum distribution curves (MDCs). Fig.~\ref{S10}f-\ref{S10}j show the photon energy dependent MDCs extracted from Fig.~\ref{S10}a-\ref{S10}e at binding energies of 0.1 eV (Fig.~\ref{S10}f), 0.3 eV (Fig.~\ref{S10}g), 0.5 eV (Fig.~\ref{S10}h), 0.7 eV (Fig.~\ref{S10}i) and 0.9 eV (Fig.~\ref{S10}j). 
The gray dashed lines in Fig.~\ref{S10}f-\ref{S10}j mark the MDCs' peak positions, which correspond to the band features marked by orange arrows. 
The negligible photon energy dependence of the MDC peak positions indicates the surface nature of these band features (marked by orange arrows in Fig.~\ref{S10}a-\ref{S10}e). By further comparing with the corresponding surface projected bands from the DFT calculations shown in Fig. 4i, we conclude that these features are SFAs.

\begin{figure*}[h]
\begin{center}
\includegraphics[width=1\columnwidth,angle=0]{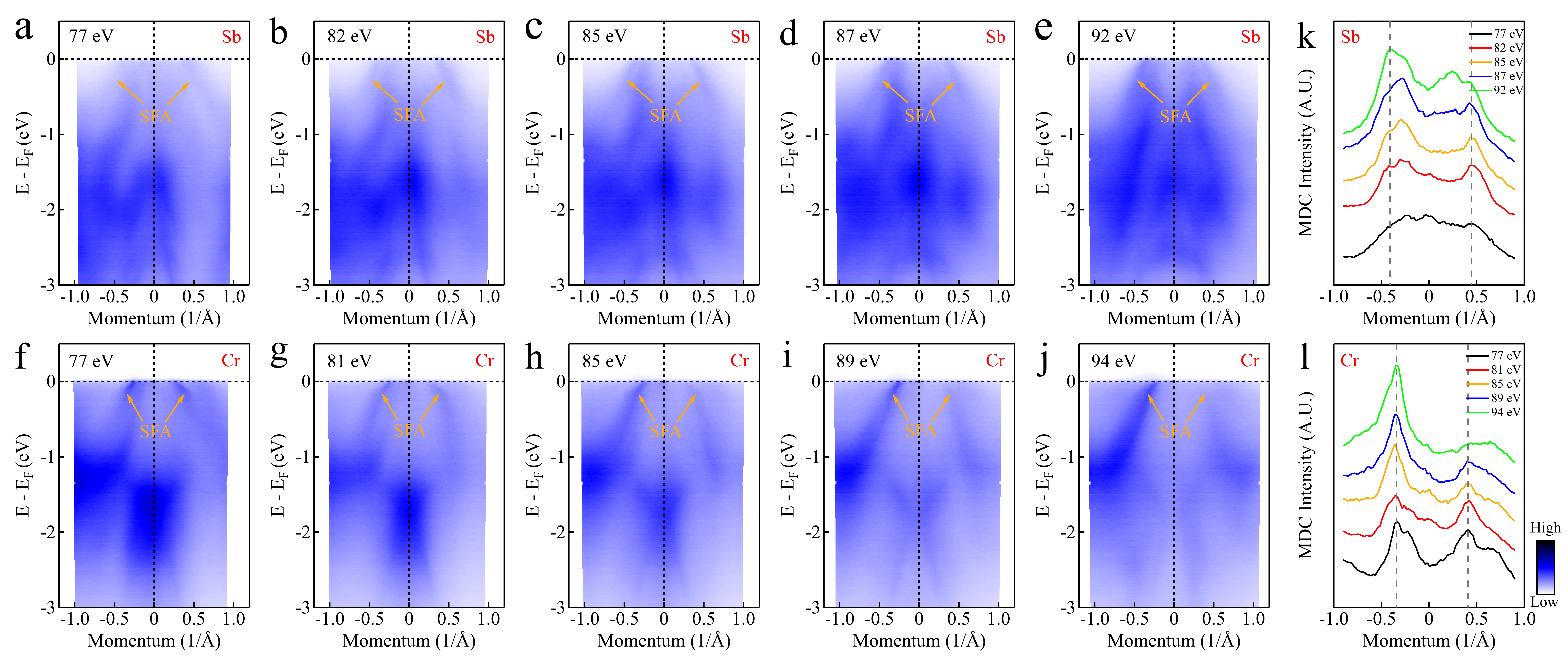}
\end{center}
\caption{\label{S11}\textbf{Photon energy dependent band dispersion along $\rm\overline{\Gamma}-\overline{K}$ on Sb and Cr terminated surfaces.} (a-e) The photon energy dependent band dispersion measurements along the $\rm\overline{\Gamma}-\overline{K}$ direction on the Sb terminated surface with photon energies of 77 ev (a), 82 eV (b), 85 eV (c), 87 eV (d) and 92 eV (e). (f-j) The photon energy dependent band dispersion measurements along the $\rm\overline{\Gamma}-\overline{K}$ direction on the Cr terminated surface with photon energies of 77 ev (f), 81 eV (g), 85 eV (h), 89 eV (i) and 94 eV (j). (k-l) The MDCs extracted from (a-e) and (f-j) at a binding energy of 0.2 eV. The MDC peaks marked by gray dashed lines in (k) and (l).
}
\end{figure*}

Fig.~\ref{S11} shows the photon energy dependent band dispersion measurements along the $\rm\overline{\Gamma}-\overline{K}$ direction on Sb (Fig.~\ref{S11}a-\ref{S11}e) and Cr (Fig.~\ref{S11}f-\ref{S11}j) terminated surfaces. 
Fig.~\ref{S11}k and \ref{S11}l show the extracted MDCs from Fig.~\ref{S11}a-\ref{S11}e and Fig.~\ref{S11}f-\ref{S11}j at a binding energy of 0.2 eV. 
The gray dashed lines in Fig.~\ref{S11}k and \ref{S11}l mark the MDC peaks corresponding to the bands marked by orange arrows in Fig.~\ref{S11}a-\ref{S11}e and Fig.~\ref{S11}f-\ref{S11}j. 
Also in this case, the negligible photon energy dependence of the MDC peaks (Fig.~\ref{S11}k and \ref{S11}l) indicate the surface nature of the bands marked by orange arrows in Fig.~\ref{S11}a-\ref{S11}e and Fig.~\ref{S11}f-\ref{S11}j. 
Comparing with the corresponding DFT calculated surface states shown in Fig. 4p and 4s, the SFA nature of the states can be determined.\\

\newpage

\section{Nodal line and drumhead surface state in CrSb without SOC}

The topological nodal line appears in systems with a band crossing and forms a closed loop. It is often constrained to a high symmetry line or plane, protected by inversion, mirror, and spin rotation symmetry \cite{TNL_review,CoS2_FM_NLS,CoS2_FM_NLS2,KNL}. 
In the absence of SOC, time-reversal symmetry ($T$), with the symmetry operator $T=\mathcal{K}$ (complex conjugation) for one spin sector, is automatically preserved, and inversion symmetry is present in the space group. The combination of time-reversal symmetry and inversion symmetry ($PT$) is thus preserved across the entire BZ. In the following, we present the $PT$ symmetry protected nodal line and its corresponding drumhead surface state. 

\begin{figure*}[h]
\begin{center}
\includegraphics[width=0.9\columnwidth,angle=0]{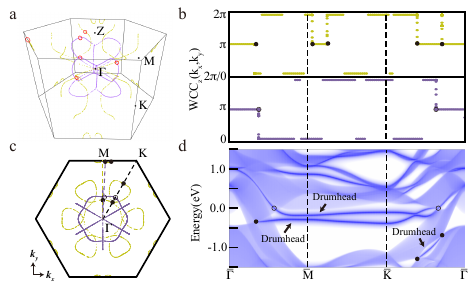}
\end{center}
\caption{\label{S12}\textbf{nodal line and surface states in the spin-up channel} (a) nodal line distribution in the 1$^{st}$ BZ where the yellow and purple dots represent the nodal lineswith occupation number 8 and 9, respectively. (b) Wannier charge center (WCC) results along the high symmetry lines $\Gamma - M - K - \Gamma$ and integrated along: $k_z$. Upper and lower panel represent the occupation number 8 and 9, respectively. (c) The top view of nodal line distribution with high symmetry lines. Open and filled circles indicate positions where the WCC jump happens in (b). (d) Surface state results on the (001) Sb terminated surface. The drumhead surface state and the corresponding projected nodal line position is labelled.
}
\end{figure*}

Around the Fermi level, we found three sets of nodal lines (see Fig.~\ref{S12}a and Fig.~\ref{S14}a-\ref{S14}c). To reveal their topological properties, we employ a Berry phase calculation along those nodal lines. As shown in Fig.~\ref{S12}a and Fig.~\ref{S14}c, all nodal lines have an integrated Berry phase $\pm \pi$ at the red circle, reflecting their nontrivial topology. 
Since the density of states above the Fermi level is relatively large, we only focus on occupation band number 8 and 9 in the following discussion. 
We then calculate the Wannier charge center (WCC) along the high symmetry lines labelled in Fig.~\ref{S12}c and the surface state calculation to determine the drumhead surface state on the (001) surface. 
As shown in Fig.~\ref{S12}b-\ref{S12}d, the change of the WCC is always associated with a crossing of the nodal line (labelled as filled and open points for occupation band number 8 and 9, respectively). 
The energies of these crossing points in Fig.~\ref{S12}b from left to right are: -0.34722, -1.51709, -1.55667, -1.27092, -6.80210 for band 8, and -0.10452, -0.00590 for band 9.
As a double check, we show the three corresponding drumhead surface states in Fig.~\ref{S12}d, all of which possess starting/ending points aligned with the WCC jump positions in Fig.~\ref{S12}b. 
We note that, due to the coincidence of the projected nodal line and $\overline{\Gamma}-\overline{M}$ path, we choose a path that slightly deviated from $\Gamma-M$ in the WCC simulations.
Despite this, the WCC result in Fig.~\ref{S12}b  matches well with the drumhead surface states in Fig.~\ref{S12}d.\\ 

\newpage

\section{SOC induced nodal line to same-spin WPs}

\begin{figure*}[h]
\begin{center}
\includegraphics[width=0.9\columnwidth,angle=0]{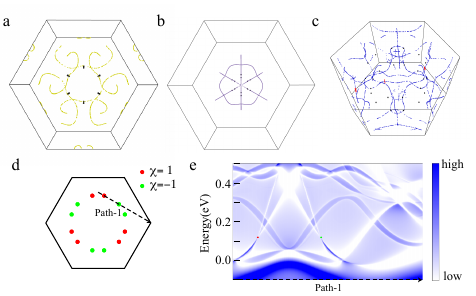}
\end{center}
\caption{\label{S14}\textbf{nodal line and same-spin WPs} (a)-(c) are the same-spin WPs distribution and the corresponding nodal linesbetween $N$ and $N+1$. The nodal linesare calculated in the spin-up channel without SOC and the WPs are the three same-spin WPs groups in Sec.~2. (c) WPs distribution of No. 3.1 with chirality denoted. The surface state on the (001) Sb terminated surface along the Path-1 denoted in (d) is presented in (e). (e) The pure surface state spectral density with the bulk state density subtracted.
}
\end{figure*}

As discussed in the last section, the nodal line exists in CrSb for both the spin-up and spin-down channel without SOC. In Fig.~\ref{S14}a-\ref{S14}c, the corresponding same-spin WPs are denoted in their SOC-free nodal lines. From Fig.~\ref{S14}a to \ref{S14}c, they are the same-spin WPs No. 3.2 and nodal line between band 8 and 9, No. 3.3 nodal line between band 9 and 10, and No. 3.1 nodal line between band 9 and 10. This SOC-induced topological nodal line to Weyl semimetal transition \cite{TNL_review} has also been reported in TaAs\cite{TaAs_NLS,TaAs_NLS2} and ZrTe\cite{ZrTe_NLS}. Here, we found two same-spin WP groups along the projected high symmetry line: $\overline{\Gamma}-\overline{M}$ that are also discussed in the main-text. These two sets of WPs also give rise to long surface states across the surface BZ. WPs No. 3.1 are buried in the bulk state and in Fig.~\ref{S14}e, clear surface state signal is shown in the pure surface state calculation results. Due to the large density of states above the Fermi level, many more WPs may be found in the conduction band region. Based on the spin-polarized Fermi arcs, the evidence of a nodal line to WPs transition, and the altermagnetic nature of CrSb with weak SOC, we conclude that the surface states in the vicinity of the Fermi level are topological.

\newpage

\section{Spin-polarized surface state and corresponding same-spin WPs}

\begin{figure*}[h]
\begin{center}
\includegraphics[width=0.9\columnwidth,angle=0]{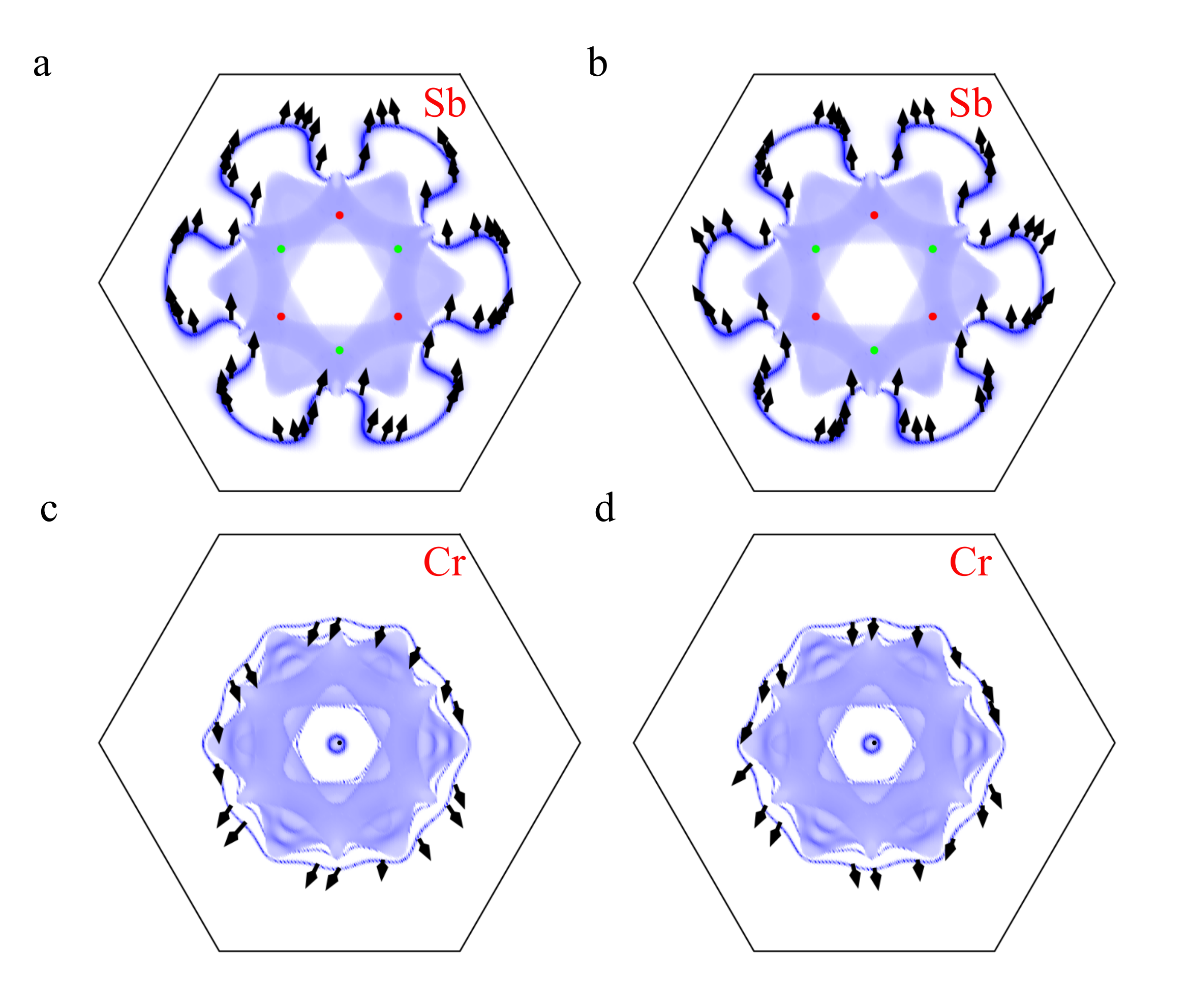}
\end{center}
\caption{\label{S13}\textbf{Surface states at $E=E_f-0.25$ eV with spin texture} (a) and (b) are surface states on the Sb terminated surface with the WPs No. 3.2 denoted; (c) and (d) are surface states on the Cr terminated surface. The corresponding spin textures are indicated by arrows where the $s_x-s_z$ and $s_y-s_z$ projections are shown in the left and right panel, respectively.
}
\end{figure*}

As the same-spin WPs can host different topological charges, the corresponding Fermi arc on the surface is naturally spin-polarized. The upper panel in Fig.~\ref{S13} shows the surface state from same-spin WPs No.\ 3.2, while the lower panel shows the surface state originating from the energy $E \approx E_f +0.3$~eV and belongs to WPs No. 3.3. On the (001) surface, two WPs with the same chirality are projected onto the same location once they are out of the $k_z=0$ plane. The total chirality $\chi = \pm 2$ of WPs No. 3.2 then results in two surface states, as shown in Fig.~\ref{S3}c-\ref{S3}d. One of these surface states disperses towards the $\Gamma$ point and another backwards. In Fig.~\ref{S13} upper panel, the Fermi arc at $E=E_f-0.25$~eV only includes one of the branches from the surface state of WPs No. 3.2. The petal-like surface states are from the drumhead surface state without SOC and half of them are now protected by the WPs with SOC. This explains the Fermi arcs in the upper panel of Fig.~\ref{S13}: the nodal line to WPs transition is driven by SOC and preserves the Fermi arcs. Since all the WPs contributing to the long Fermi arcs are of the same-spin type and their Fermi arcs dispersions are surface dependent, their corresponding Fermi arcs are spin-polarized and also separated on different terminations.\\ 

\newpage